\documentclass[sn-mathphys]{sn-jnl}

\jyear{2021}%

\theoremstyle{thmstyleone}%
\theoremstyle{thmstyletwo}%

\theoremstyle{thmstylethree}%

\begin{document}

\title[Article Title]{MFV approach to robust estimate of neutron lifetime}


\author*[1,2,3]{\fnm{Jiang} \sur{Zhang}}\email{zhangphysics@126.com}

\author[1]{\fnm{Sen} \sur{Zhang}} 
\author[2]{Zhen-Rong Zhang}
\author[1]{Pu Zhang}
\author[1]{Wen-Bin Li}
\author*[1]{Yan Hong} \email{hongyandemail@163.com}

\affil[1]{Physics Experimental Center, Department of Mathematics and Physics, Hebei GEO University, Shijiazhuang 050016, China}
\affil[2]{School of Basic Sciences, Tianjin Agricultural University, Tianjin 300384, China }
\affil[3]{Hebei Key Laboratory of Optoelectronic Information and Geo-detection Technology, Hebei
	GEO University, China}


\abstract{Aiming at evaluating the lifetime of the neutron, we introduce a novel statistical method to analyse the updated compilation of precise measurements including the 2022 dataset of Particle Data Group (PDG). Based on the minimization for the information loss principle, unlike the median statistics method, we apply the most frequent value (MFV) procedure to estimate the neutron lifetime, irrespective of the Gaussian or non-Gaussian distributions. Providing a more robust way, the calculated result of the MFV is $\tau_n=881.16^{+2.25}_{-2.35}$ s with statistical bootstrap errors, while the result of median statistics is $\tau_n=881.5^{+5.5}_{-3}$ s according to the binomial distribution. Using the different central estimates, we also construct the error distributions of neutron lifetime measurements and find the non-Gaussianity, which is still meaningful.}

\keywords{neutron, lifetime, data analysis}



\maketitle

\section{Introduction}\label{sec1}

It is well-known that the decay of a free neutron related to the weak
interaction is the most important $\beta$-decay process
producing a proton, an electron, and an antineutrino with a lifetime of about 880 seconds.
Furthermore, the neutron lifetime plays a fundamental role not only in particle physics but also astrophysics, such as solar physics and cosmology. 
Obviously, a more accurate value of neutron lifetime will improve our understanding of related fields \cite{Wietfeldt2011, Wietfeldt2018}.

On account of the great significance of the $\beta$-decay of the free neutron, there are three different experimental methods to measure the free neutron lifetime over the last 60 years \cite{Wietfeldt2011}. For the first method, the neutron lifetime in 'beam' experiments can be evaluated on the neutron decay rate from the number of decay particles \cite{Yue2013}. For the second method, the neutron lifetime in 'bottle' experiments can be determined by measuring the remaining number of ultra-cold neutrons stored in the container \cite{Huffman2000,Byrne2019,Serebrov2018}. Because the magnetic trap technique can overcome the disadvantages of the traditional material wall, recently several studies have utilized a magneto-gravitational trap technique to determine a neutron lifetime with increasing precision\cite{Pattie2018, Gonzalez2021}. The third method applies the space-based technique to determine the  neutron lifetime using the neutron spectrometer on NASA's lunar prospector mission\cite{Wilson2021}. It is noteworthy that over the past decades, many more precise measurements of neutron lifetime are obtained, which prompted the research on theoretical physics. However, there is still a significant discrepancy ($\sim4\sigma$\cite{Gonzalez2021,Fornal2018}) among the above methods, which has long been a puzzling problem involving particle physics, nuclear physics, and astrophysics.

The conventional median technique is one of the most widely used statistical procedures to estimate the characteristics of different observational quantities in many statistical applications because it is not sensitive to outliers \cite{Gott2001}. There have been a number of known  examples of non-Gaussian data \cite{Bailey2017} to apply the median statistics, involving Hubble constant \cite{Chen2003a,Chen2011}, $^7$Li abundance \cite{Crandall2015MPLA,Zhang2017}, LMC and SMC distance  \cite{Crandall2015ApJ}, deuterium abundance and spatial curvature constraints \cite{Penton2018}, the distance to the Galactic center\cite{Camarillo2018PASP}, galactic rotational velocity \cite{Camarillo2018Apss,Rajan2018} and  Newton's constants\cite{Bethapudi2017}. Recently, Rajan \& Desai (2020) \cite{Rajan2020} investigated the measurements of neutron lifetime tabulated by Tanabashi et al. from Particle Data Group (PDG, hereafter, the 2019 edition) \cite{Tanabashi2018}  based on a meta analysis. Furthermore, the results demonstrated that the error distribution of neutron lifetime measurements is not a Gaussian feature. Especially, the contributions of unidentified systematic effects and uncertainties are proposed to explain the anomalous discrepancy between different methods of neutron lifetime, which might imply the existence of new physics.

Obviously, from the standpoint of robust statistics, it is very important to estimate the true value of neutron lifetime considering the holistic characteristic of all observations. Despite a great deal of research on data analysis and uncertainty in physics and technology over the past several decades, the current status of discrepancy among different measurements still needs a novel statistical method to enhance the robustness of the model, which has motivated us to apply the MFV procedure \cite{Steiner1988, Steiner1991,Steiner1997,Steiner2001} to reanalyze this tension. Besides the conventional statistical algorithms, as a robust and resistant procedure, the MFV statistics have been used widely to seek the robust estimate in different natural science problems\cite{Kemp2006, Szucs2006,Szegedi2013,Szegedi2014,Szabo2018}, such as the lithium abundances problem and Hubble constant tension \cite{Zhang2017,Zhang2018}.

In the next section we describe the latest compilation of neutron lifetime in PDG. Section 3 utilizes the MFV method to estimate the neutron lifetime value. To compare with past traditional results, we also calculate the confidence intervals and illustrate its advantage of MFV. In Section 4 we describe our analysis of error distributions around the different central estimates (the weighted mean, median, and MFV). Conclusions are given in Section 5.

\section{Neutron lifetime data}\label{sec2}

Up to now, many measured values of neutron lifetime using different techniques have been published, e.g., Ezhov et al. (2018) \cite{Ezhov2018} with $\tau_n=878.3\pm1.6_{stat}\pm1.0_{sys}$ s, Serebrov et al. (2018) \cite{Serebrov2018} with $\tau_n=881.5\pm0.7_{stat}\pm0.6_{sys}$ s, Pattie et al. (2018) \cite{Pattie2018} with $\tau_n=877.7\pm0.7_{stat}\pm0.4/-0.2_{sys}$ s, Leung et al. (2016) \cite{Leung2016} with $\tau_n=887\pm39$ s, Arzumanov et al. (2015)\cite{Arzumanov2015} with $\tau_n=880.2\pm1.2$ s, Yue et al. (2013)\cite{Yue2013} with $\tau_n=887.7\pm1.2_{stat}\pm1.9_{sys}$, Steyerl et al. (2012) \cite{Steyerl2012} with $\tau_n=882.5\pm2.1$ s, Pichlmaier et al. (2010)\cite{Pichlmaier2010} with $\tau_n=880.7\pm1.3\pm1.2$ (also see Rajan \& Desai 2020 \cite{Rajan2020} and Tanabashi et al. 2018 \cite{Tanabashi2018}). In our analysis, we use the updated compilation from Rajan \& Desai (2020) \cite{Rajan2020} and the latest results of PDG, including 	
Gonzalez et al. 2021 \cite{Gonzalez2021} and Wilson et al. 2021 \cite{Wilson2021} (see https://pdglive.lbl.gov/DataBlock.action?node=S017T, the 2022 edition of PDG\cite{PDG22}). Moreover, we adopt descriptive statistics methods to analyze the used measurements of neutron lifetime and plot the histogram of the number of all data, as shown in Fig. 1. Following graphical depictions of the observed data as a function of publication data\cite{de_Grijs2014AJa,de_Grijs2020ApJS248},
Fig. 2 illustrates a summary of neutron lifetime experimental results used for analysis since 1972, including measurements listed in/not\_in PDG and results unused in PDG following the taxonomic approach of Rajan \& Desai (2020)\cite{Rajan2020}. Moreover, the marginal panels associated with the main panel show the distribution of neutron lifetime contributing to the published years. As summarized in Fig. 3, we give a graphical representation of neutron lifetime data based on beam, bottle, and space-based methods  provided by different authors. Then it becomes apparent that the highly frequent data focus on bottle experiments, and many more space-based measurements are required\cite{Wilson2021}.  Besides, it is  possible to constrain neutron lifetime from cosmological $Y_p$ values and Big Bang nucleosynthesis\cite{Salvati2016}, which is model-dependent. Following Rajan \& Desai (2020)\cite{Rajan2020}, we also do not consider cosmological parameter estimation in our calculation. Finally, we present all neutron lifetime measurements from PDG and Rajan \& Desai (2020)  in Table 1.

\begin{table}[htbp]
	\centering
	\caption{21 neutron lifetime measurements from PDG. The taxonomic approach of Not\_in\_PDG, PDG and PDG\_not\_used for analysis is similar to Rajan \& Desai (2020).}
	\begin{tabular}{ccccc}
		\toprule
		\multicolumn{1}{l}{Neutron lifetime (secs)} & Reference & Type  & Comment &  \\
		\midrule
		$878.3\pm1.6\pm1.0$ & Ezhov 2018\cite{Ezhov2018} & Bottle & PDG   &  \\
		$881.5\pm0.7\pm0.6$ & Serebrov 2018\cite{Serebrov2018} & Bottle & PDG   &  \\
		$877.7\pm0.7+0.4/-0.2$ & Pattie 2018\cite{Pattie2018} & Bottle & PDG   &  \\
		$887\pm39$   & Leung 2016\cite{Leung2016} & Bottle & Not in PDG &  \\
		$880.2\pm1.2$ & Arzumanov 2015\cite{Arzumanov2015} & Bottle & PDG   &  \\
		$887.7\pm1.2\pm1.9$ & Yue 2013\cite{Yue2013} & Beam  & PDG   &  \\
		$882.5\pm1.4\pm1.5$ & Steyerl 2012\cite{Steyerl2012} & Bottle & PDG   &  \\
		$880.7\pm1.3\pm1.2$ & Pichlmaier 2010\cite{Pichlmaier2010} & Bottle & PDG   &  \\
		$878.5\pm0.7\pm0.3$ & Serebrov 2005\cite{Serebrov2005} & Bottle & PDG   &  \\
		$889.2\pm3.0\pm3.8$ & Byrne 1996\cite{Byrne1996} & Beam  & PDG   &  \\
		$882.6\pm2.7$ & Mampe 1993\cite{Mampe1993} & Bottle & PDG   &  \\
		$888.4\pm2.9$ & Alfikmenov 1990\cite{Alfimenkov1990} & Bottle & PDG\_not\_used &  \\
		$878\pm27\pm14$   & Kossakowski 1989\cite{Kossakowski1989} & Beam  & PDG\_not\_used &  \\
		$877\pm10$   & Paul 1989\cite{Paul1989} & Bottle & PDG\_not\_used &  \\
		$876\pm10\pm19$   & Last 1988\cite{Last1988} & Beam  & PDG\_not\_used &  \\
		$891\pm9$   & Spivak 1988\cite{Spivak1988}  & Beam  & PDG\_not\_used &  \\
		$903\pm13$   & Kosvintsev 1986\cite{Kosvintsev1986} & Bottle & PDG\_not\_used &  \\
		$875\pm95$   & Kosvintsev 1980\cite{kosvintsev1980} & Bottle & PDG\_not\_used &  \\
		$918\pm14$   & Christensen 1972\cite{Christensen1972} & Beam  & PDG\_not\_used &  \\
		$877.75\pm0.28+0.22/-0.16$ &Gonzalez 2021\cite{Gonzalez2021} & Bottle & PDG   &  \\
		$887\pm14+7/-3$   & Wilson 2021\cite{Wilson2021} & Space & PDG\_not\_used &  \\
		\bottomrule
	\end{tabular}%
	\label{tab:addlabel}%
\end{table}%

\vskip 4mm
\centerline{\includegraphics[width=0.9\textwidth]{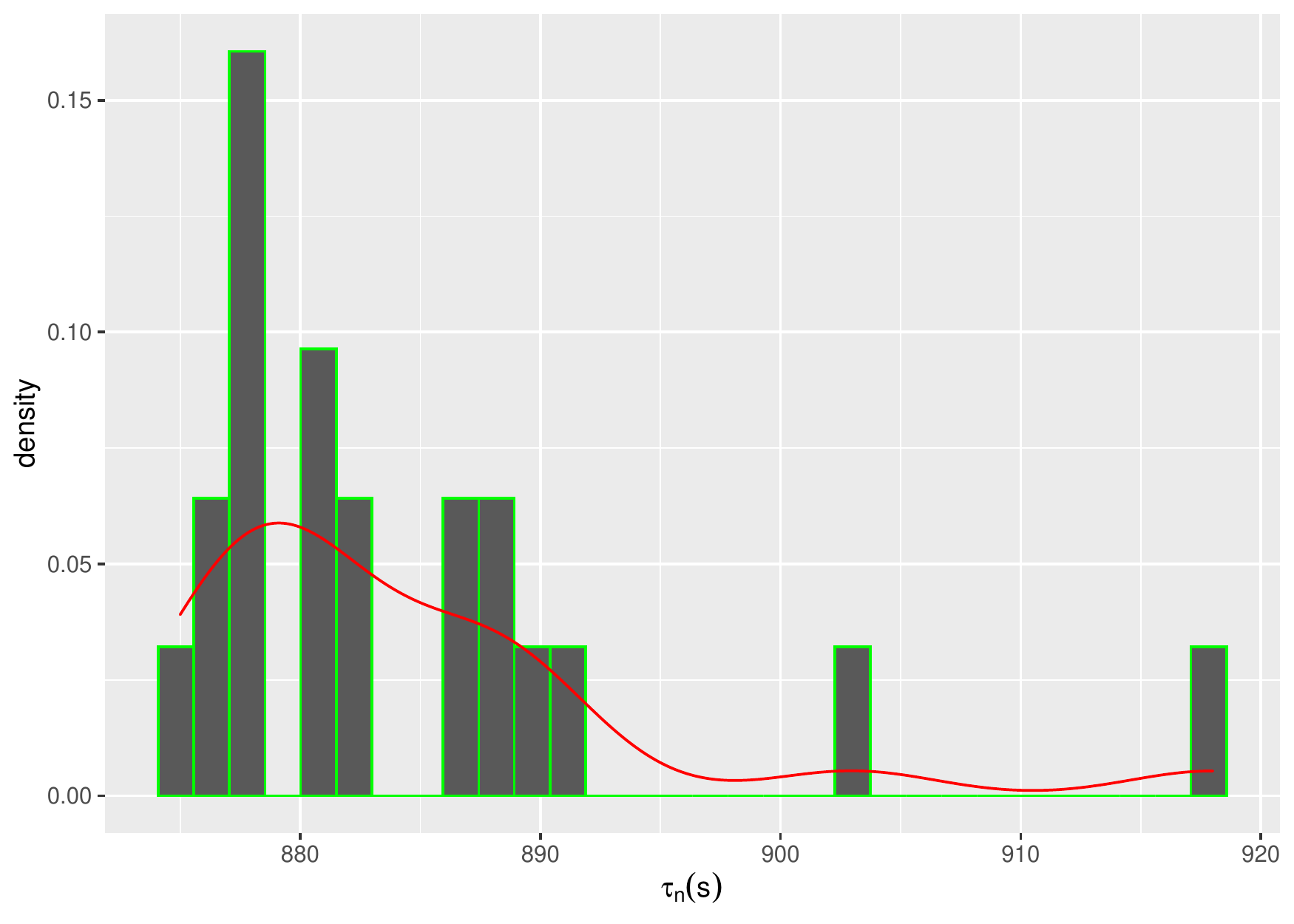}}
\vskip 2mm

\centerline{\footnotesize \begin{tabular}{p{10cm}}\bf Fig.\,1. \rm
		Histogram and probability density (red line) of neutron lifetime measurements. [A colour version of this figure is available in the online version.]
\end{tabular}}

\vskip 0.5\baselineskip

\centerline{\includegraphics[width=0.9\textwidth,height=0.4\textheight]{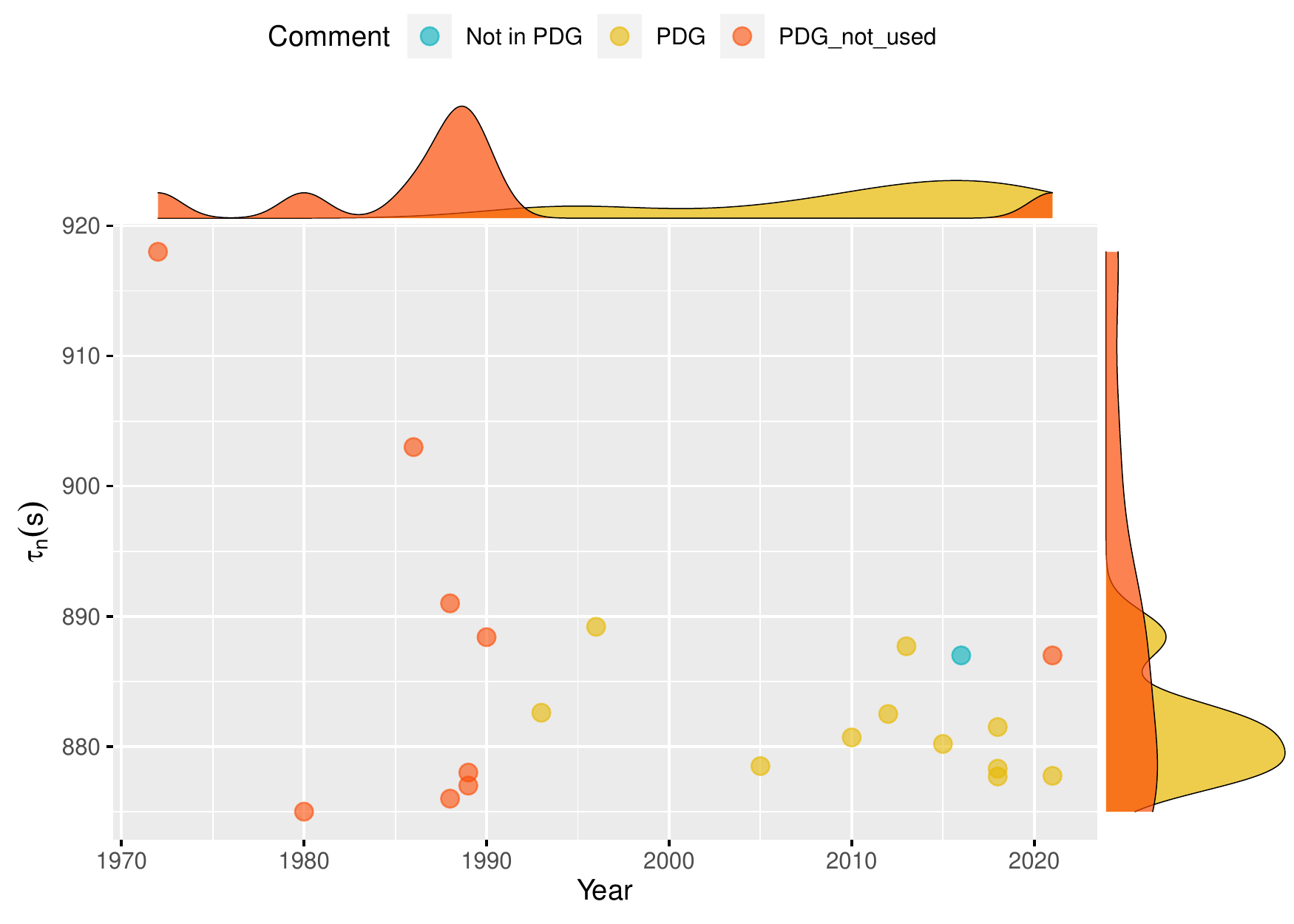}}
\vskip 2mm
\centerline{\footnotesize \begin{tabular}{p{10cm}}\bf Fig.\,2. \rm
		Published neutron lifetime as a function of publication date. The taxonomic approach of Not\_in\_PDG, PDG and PDG\_not\_used for analysis is similar to Rajan \& Desai (2020).  [A colour version of this figure is available in the online version.]
\end{tabular}}
\vskip 0.5\baselineskip

\centerline{\includegraphics[width=0.9\textwidth,height=0.4\textheight]{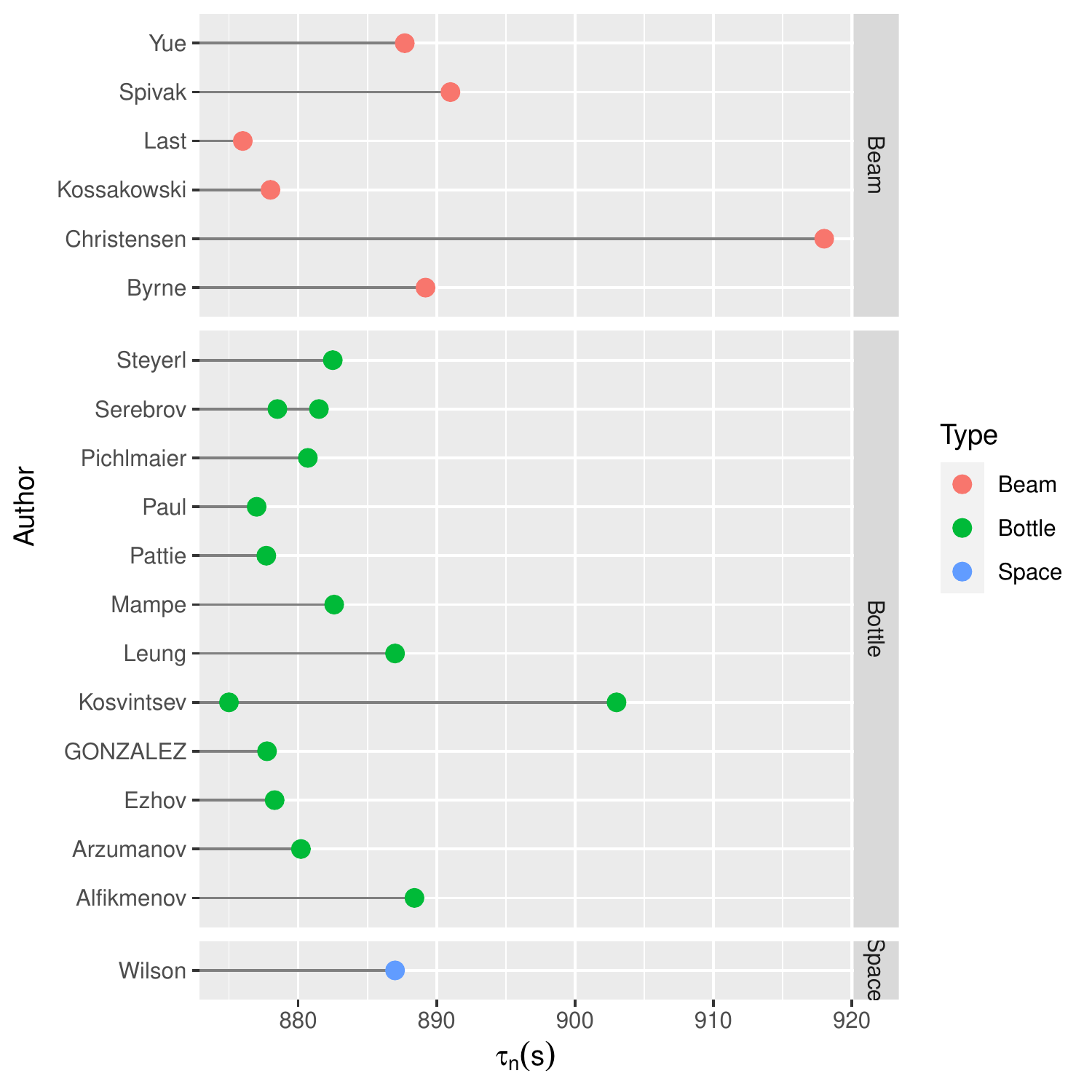}}
\vskip 2mm
\centerline{\footnotesize \begin{tabular}{p{10cm}}\bf Fig.\,3. \rm
		A summary of the experimental results used for the analysis from different authors. [A colour version of this figure is available in the online version.]
\end{tabular}}
\vskip 0.5\baselineskip

\section{Analysis}\label{sec3}

In nuclear physics fields, the estimation of neutron lifetime from the observed measurements is one of the vital problems \cite{Rajan2020}. For decades, a considerable amount of statistical methods have been widely used, such as 
median statistics \cite{Chen2011, Gott2001, Crandall2015b}, maximum likelihood estimation, and Bayesian statistics\cite{Erler2020}. Naturally, there is no motive for researchers to permanently suppose that the prior distribution of a physical quantity is always normal. In many cases, Gaussian distribution is not the only way even considering the larger sample. For example,  it is the most common situation that if the dataset produces from Cauchy distribution, the sampling distribution is non-Gaussian, regardless of the size of the sample space.

In general, $\chi^{2}$ analyses, least square method (LSM) and  median statistics are utilized to obtain the key information from the measurement data \cite{Gott2001,Rajan2020}. Moreover, provided that the feature of prior distributions is non-Gaussian, differences in the prior distributions need to be taken into  account in the calculation of more precise results. On the other hand, as is known to all that George E. P. Box gave the  famous quote ``all models are wrong, but some are useful''. 
Every computational statistical model may have some additional potential hypotheses. Obviously,  we should ruminate the normality of distribution, including the error distribution and prior distribution before statistical computation \cite{Steiner1997,Zhang2017,Zhang2018}. In addition, the heteroscedastic analysis is always a big problem for the conventional procedures. Thus, it is necessary to uncover the significant details about distributions of neutron lifetime measurements, shown in Fig. 4 and Fig. 5. These figures present the violin plot and kernel density plot for different methods, which is especially useful for analyzing summary and descriptive statistics. Furthermore, we can directly compare the median and MFV values, and intuitively see the distributions of different subgroups implying the non-Gaussian errors in neutron lifetime data of different approaches.

On the base of central limit theorem, the distribution of measurement data should be normal in most cases. Even so, the error distributions of measurement data  are still probably non-Gaussian. For instance, it is possible that the measurement data may not originate from a random sample of independent, identically distributed random variables, which should be proved. 
Another non-negligible point is the heavy-tailed problem of the observed distribution. Essentially, no one can be accurately aware whether or not the normality feature is intrinsic for the measurement data \cite{Bailey2017, Chen2003b, Crandall2015b, Zhang2017, Singh2016}.  In short, these difficulties also motivate us to apply the novel robust MFV method to evaluate the true value of neutron lifetime.

Because of the ideal condition of pure mathematics, it is on the ground of expediency that some plausible hypotheses have to be accepted temporarily. According to the measurement data of neutron lifetime, physicists expect to estimate accurately the real value from different prior distributions. For the sake of a better implementation of this purpose,  Steiner \cite{Steiner1991,Steiner1997} proposed a more robust statistical algorithm --- MFV method, based on the minimization for the information loss principle. Nearly regardless of considering the normality of the prior distribution, the MFV is not only highly robust efficiency but also dispose of the deficiency, for example, high sensitivity for outliers of some data\cite{Zhang2017,Zhang2018}.

In order to elucidate the effect of prior distribution and error distributions, we utilize the MFV procedure to assess the characteristics of datasets of neutron lifetime.
Unlike the traditional methods such as Maximum likelihood estimation or LSM, Steiner put forward the maximum reciprocals principle,
\begin{eqnarray}
\label{eq1}
\sum_{i} \frac{1}{X_i^2+S^2}=max,
\end{eqnarray}
where $X_i$ is the residuals or deviations, i.e., $X_i=T_i^{measured}-T_i^{computed}$, and S is the parameter of scale characterizing the measurement error, denoted by $\varepsilon$ (called dihesion).
According to the minimization of the information divergence (relative entropy) demonstrating the measure of information loss \cite{Huber1981,Steiner1991,Steiner1997}, Steiner suggested the MFV method and the scaling factor $\varepsilon$, i.e., dihesion, for the sake of evaluating the parameter of scale to some extent to reduce the information loss. Furthermore,   using the iteratively re-weighted least squares procedure, we can calculate the MFV and the dihesion via iterations\cite{Steiner1997,Steiner2001,Szucs2006}. Especially, Steiner had proved the MFV procedure has the advantages of resistance and robustness\cite{Steiner1991,Steiner1997}.  After the (j+1)-th step of the MFV procedure, the relative equation of iterations for the most frequent value $M$ is as follows:
\begin{eqnarray}
\label{eq1}
M_{j+1}= \frac{ \sum_{i=1}^{n} \frac{\varepsilon_j^2x_i}{\varepsilon_j^2+(x_i-M_j)^2}} {\sum_{i=1}^{n}\frac{\varepsilon_j^2}{\varepsilon_j^2+(x_i-M_j)^2}},
\end{eqnarray}
where $x_i$ is a series of the measurements and the dihesion $\varepsilon_j$ can be calculated by
\begin{eqnarray}
\label{eq1}
\varepsilon_{k+1}^2= \frac{3 \sum_{i=1}^{n} \frac{\varepsilon_k^4(x_i-M_j)^2}{[\varepsilon_k^2+(x_i-M_j)^2]^2}} {\sum_{i=1}^{n}\frac{\varepsilon_k^4}{[\varepsilon_k^2+(x_i-M_j)^2]^2}}.
\end{eqnarray}
Where the iterative initial value $M_0$ is chosen as the average value of the measurements, and the initial value of $\varepsilon$ is obtained as
\begin{eqnarray}
\label{eq1}
\varepsilon_0=\frac{\sqrt{3}}{2}(x_{max}-x_{min}).
\end{eqnarray}
In addition, we take the fixed threshold criterion to restrain the precision in iterations. Through a series of iterations, the most frequent value $M$ and dihesion $\varepsilon$ can be obtained  when the dihesion is smaller than threshold value (e.g. 10$^{-5}$). Apparently, the dihesion $\varepsilon$ is not like the standard deviation in LSM, which is sensitive to the outliers. 
Using all data listed in column 1 of Table 1,
we obtain the result of MFV calculations is $\tau_n =881.16$ s, which agrees with the recent experimental data. Similarly, the calculated MFV using beam and bottle-based measurements are 888.83 s and 879.85 s shown in Fig. 4, respectively.  

According to the fact that the experimental data reflect the nature of physical quantities, some statistics can be used to characterize the data, such as the MFV of the sample. To estimate the uncertainty of physical quantities, the bootstrap method is one of the most effective methods and is significant to assess the rationality of the calculated results\cite{Efron1994}. The basic process of the bootstrap method to calculate the confidence interval is as follows. Suppose that the experimental data set of neutron lifetime is ($t_1$,...,$t_i$) chosen from the distribution of true values of neutron lifetime with the corresponding statistic $\theta$($t_1$,...,$t_i$), i.e. MFV. First, we produce a bootstrap sample  ($t^*_1$,...,$t^*_i$) from the initial experimental data with replacement. Next, the important statistic, i.e. MFV, is used for the bootstrap sample. Finally, repeating this process B times (usually 1000-3000 times) generates the distribution of the MFV. Therefore, these distributions can be used to evaluate confidence intervals (usually 68.27\% or 95.45\%) of the MFV for different measuring methods of neutron lifetime. The 68.27\% confidence interval for all measurements is [878.81, 883.41], involving statistical bootstrap errors, while the 95.45\% confidence interval for all data is [877.72, 885.61].

The second technique to estimate confidence intervals is the median statistics\cite{Gott2001,Camarillo2018PASP}. Based on the binomial test and estimation in non-parametric statistics\cite{Conover1999}, the probability of the median between values $x^{(r)}$ and $x^{(s)}$ is  
\begin{eqnarray}
\label{eq1}
p(x^{(r)}\leqslant med \leqslant x^{(s)}) = p(med \leqslant x^{(s)}) - p(med < x^{(r)})
=\sum_{i=r}^{s} \binom{n}{i}/2^n,
\end{eqnarray}
where $x^{(i)}$ is the order statistic. By application of this formula, the median for all measurements is 881.5 and the calculated 68.27\% confidence interval is [878.5, 887.0],  while the 95.45\% confidence interval for all data is [878.0  887.7].

\centerline{\includegraphics[width=0.9\textwidth]{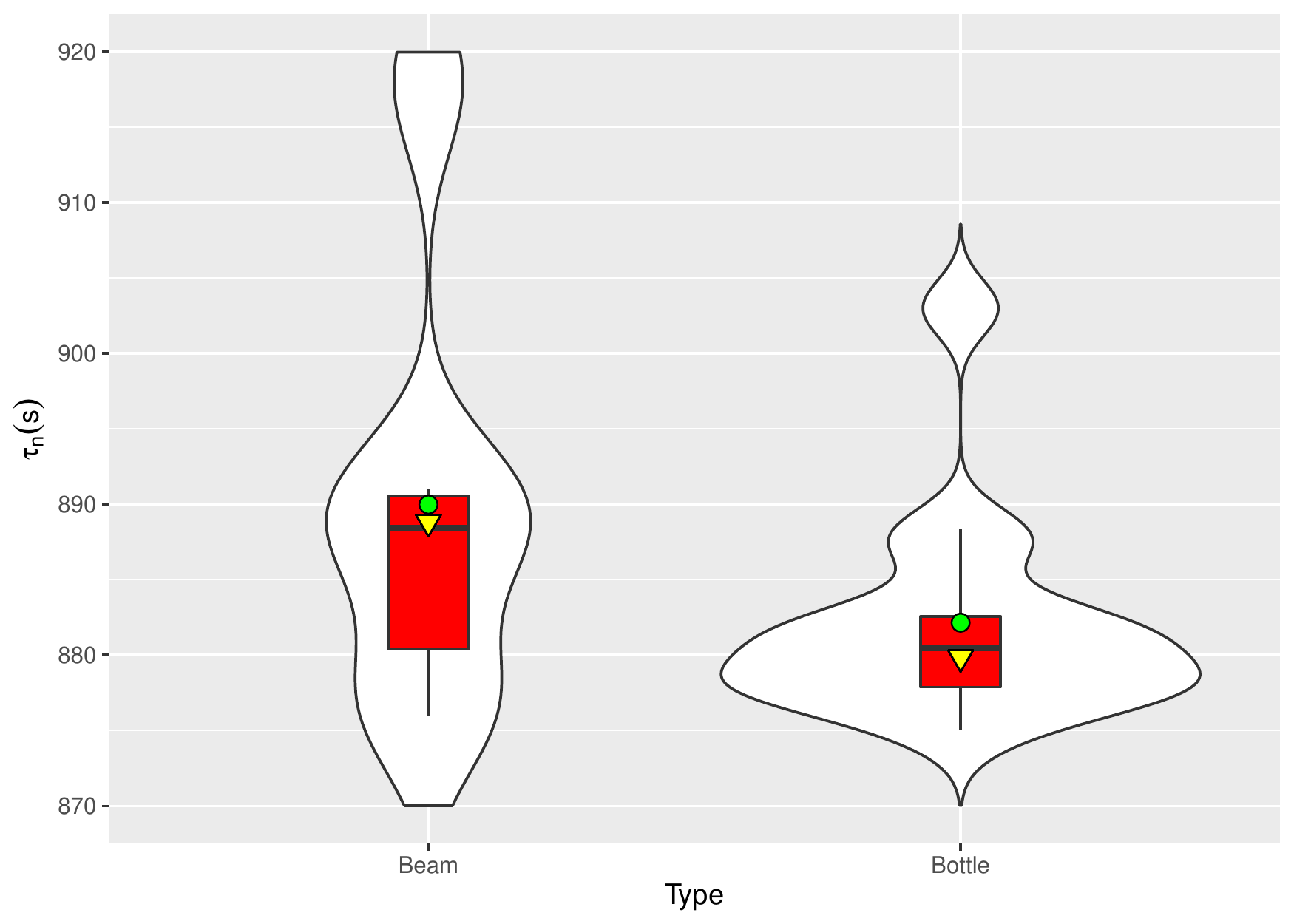}}
\vskip 2mm
\centerline{\footnotesize \begin{tabular}{p{10cm}}\bf Fig.\,4. \rm
		Violin plot of neutron lifetime measurements. The MFV and average values from different methods are indicated by the inverted triangles and circles, respectively, while the horizontal solid lines in boxes illustrate the median.  [A colour version of this figure is available in the online version.]
\end{tabular}}
\vskip 0.5\baselineskip

\centerline{\includegraphics[width=0.9\textwidth]{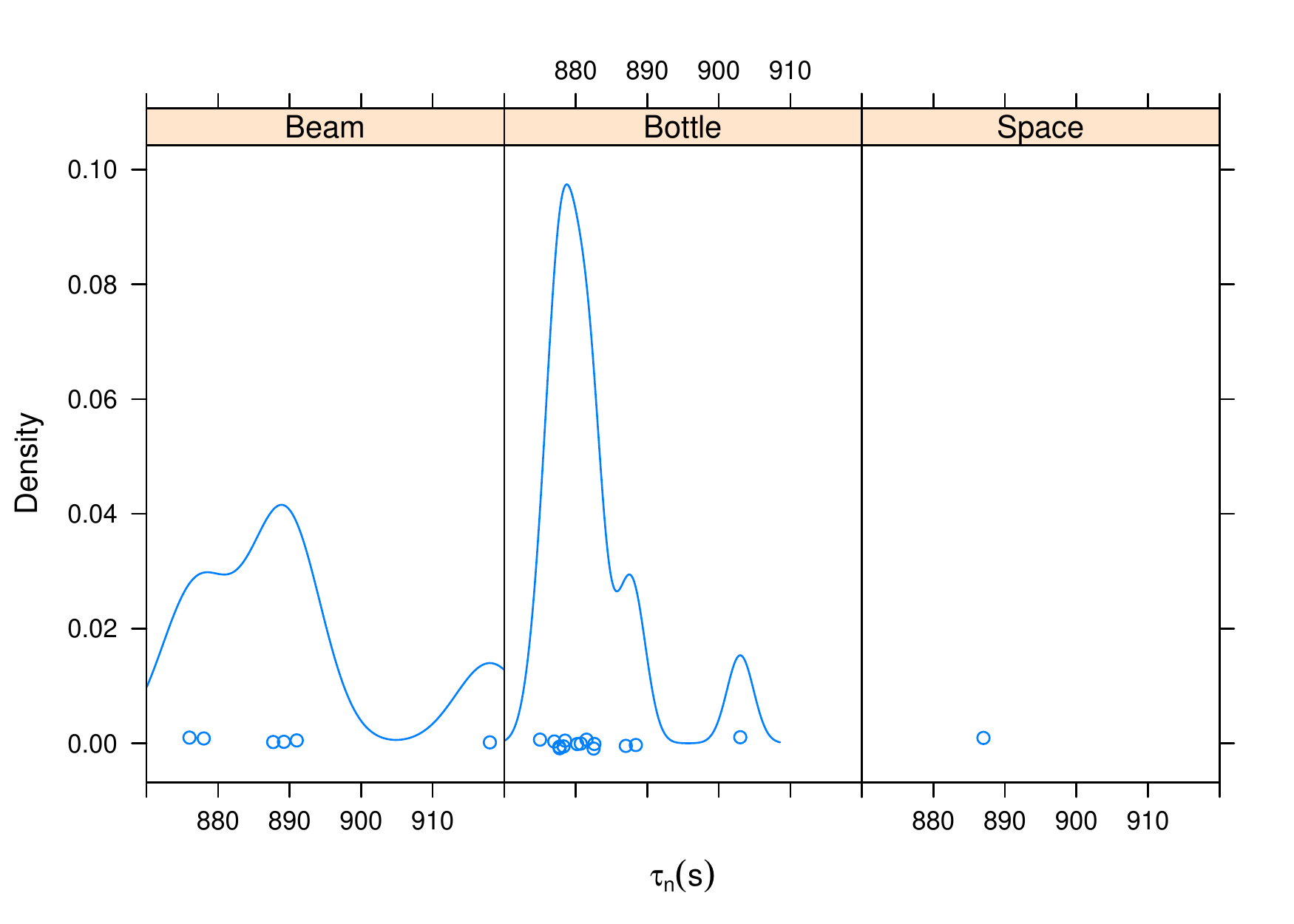}}
\vskip 2mm
\centerline{\footnotesize \begin{tabular}{p{10cm}}\bf Fig.\,5. \rm
		Density plot of neutron lifetime measurements from different methods. Circles represent the observed values. [A colour version of this figure is available in the online version.]
\end{tabular}}
\vskip 0.5\baselineskip

\section{Error Distribution For PDG Dataset}\label{sec4}

It is worthwhile to investigate the distributions of deviations of the neutron lifetime measurements mentioned in PDG 2022 from the central estimate. Among all the measured data, asymmetric systematic uncertainties should be treated in a clear and consistent manner. Especially, the asymmetric errors of $877.7\pm0.7+0.4/-0.2$\cite{Pattie2018}, and $877.75\pm0.28+0.22/-0.16$\cite{Gonzalez2021} are widely used. Nevertheless, the effect of asymmetric uncertainties are difficult to assess exactly\cite{Barlow2003, Barlow2004}. There are several ways to evaluate the asymmetric errors, including maximum absolute error, average error, piecewise linear error, quadratic error, and Fechner distribution methods\cite{Barlow2003, Barlow2004, Lista2017, Audi2017, Possolo2019, Rajan2020}. Based on these methods, we have calculated the variance, and find a tiny dispersion using the central value to be in the middle between the upper and lower uncertainties. For example as proposed by Audi et al. (2017, Appendix A)\cite{Audi2017}, we apply the formula $Var=(1-2/\pi)(a-b)^2+a*b$ to obtain $\sigma=0.307$ for Pattie et al. (2018)\cite{Pattie2018}, while the average value of asymmetric systematic errors is 0.3. Furthermore, for experimental neutron lifetime measurements, the symmetrization of asymmetric uncertainties lead to a tiny bias (generally very small in relative error) in the estimate of combined values, which can be neglected. Therefore, for simplicity, here we choose the measured midpoint and average value of asymmetric uncertainties motivated by Audi et al.\cite{Audi2012} and Barlow\cite{Barlow2019}.

There are three statistical central estimates: weighted mean\cite{Podariu2001}, median, and MFV. The weighted mean is given by 
\begin{eqnarray}
\label{eq1}
T_{wm}=\frac{\sum_{i=1}^{N}T_i/\sigma_i^2}{\sum_{i=1}^{N} 1/\sigma_i^2},
\end{eqnarray}
where $T_i$ is the measurement of neutron lifetime and $\sigma_i$ is the one standard deviation error, i.e. the quadrature sum of the statistical and systematic uncertainties. The weighted mean standard deviation is 
\begin{eqnarray}
\label{eq1}
\sigma_{wm}=\frac{1}{ \sqrt{\sum_{i=1}^{N} 1/\sigma_i^2}}.
\end{eqnarray}
The goodness of fit $\chi^2$ is
\begin{eqnarray}
\label{eq1}
\chi^2=\frac{1}{N-1} \sum_{i=1}^{N}{(T_i-T_{wm})^2}/{\sigma_i^2}.
\end{eqnarray}
The number of standard deviations that $\chi$ deviates from unity \cite{Farooq2013PLB,Crandall2014,Crandall2015ApJ} is described by 
\begin{eqnarray}
\label{eq1}
N_\sigma= \mid\chi-1\mid\sqrt{2(N-1)}.
\end{eqnarray}

We can also use the median and MFV statistics approaches to construct the error distributions. Similar to the median statistics to hypothesize statistical independence of all measurements, the MFV statistics does not utilize the individual measurement uncertainties leading to a wider interval of errors on the central value in comparison to the weighted mean technique. Under the conditions of the specified central estimate of all measurements, the error distribution associated with standard deviations \cite{Crandall2015ApJ,Camarillo2018PASP} is defined as
\begin{eqnarray}
\label{eq1}
N_{\sigma_i}= \frac{T_i- T_{CE}}{\sqrt{\sigma_i^2+\sigma_{CE}^2}},
\end{eqnarray}
where $T_{CE}$ is the central estimate of neutron lifetime measurements, either the median $T_{med}$ or MFV $T_{MFV}$, and $\sigma_{CE}$ is the uncertainty of $T_{CE}$, either $\sigma_{med}$ or $\sigma_{MFV}$. These different combinations of central estimates and uncertainties are given by 
\begin{eqnarray}
\label{eq1}
N^{wm}_{\sigma_i}= \frac{T_i- T_{wm}}{\sqrt{\sigma_i^2+\sigma_{wm}^2}},
\end{eqnarray}
\begin{eqnarray}
\label{eq1}
N^{med}_{\sigma_i}= \frac{T_i- T_{med}}{\sqrt{\sigma_i^2+\sigma_{med}^2}},
\end{eqnarray}
\begin{eqnarray}
\label{eq1}
N^{MFV}_{\sigma_i}= \frac{T_i- T_{MFV}}{\sqrt{\sigma_i^2+\sigma_{MFV}^2}}.
\end{eqnarray}

\begin{center}
	\centerline{\includegraphics[width=1\textwidth]{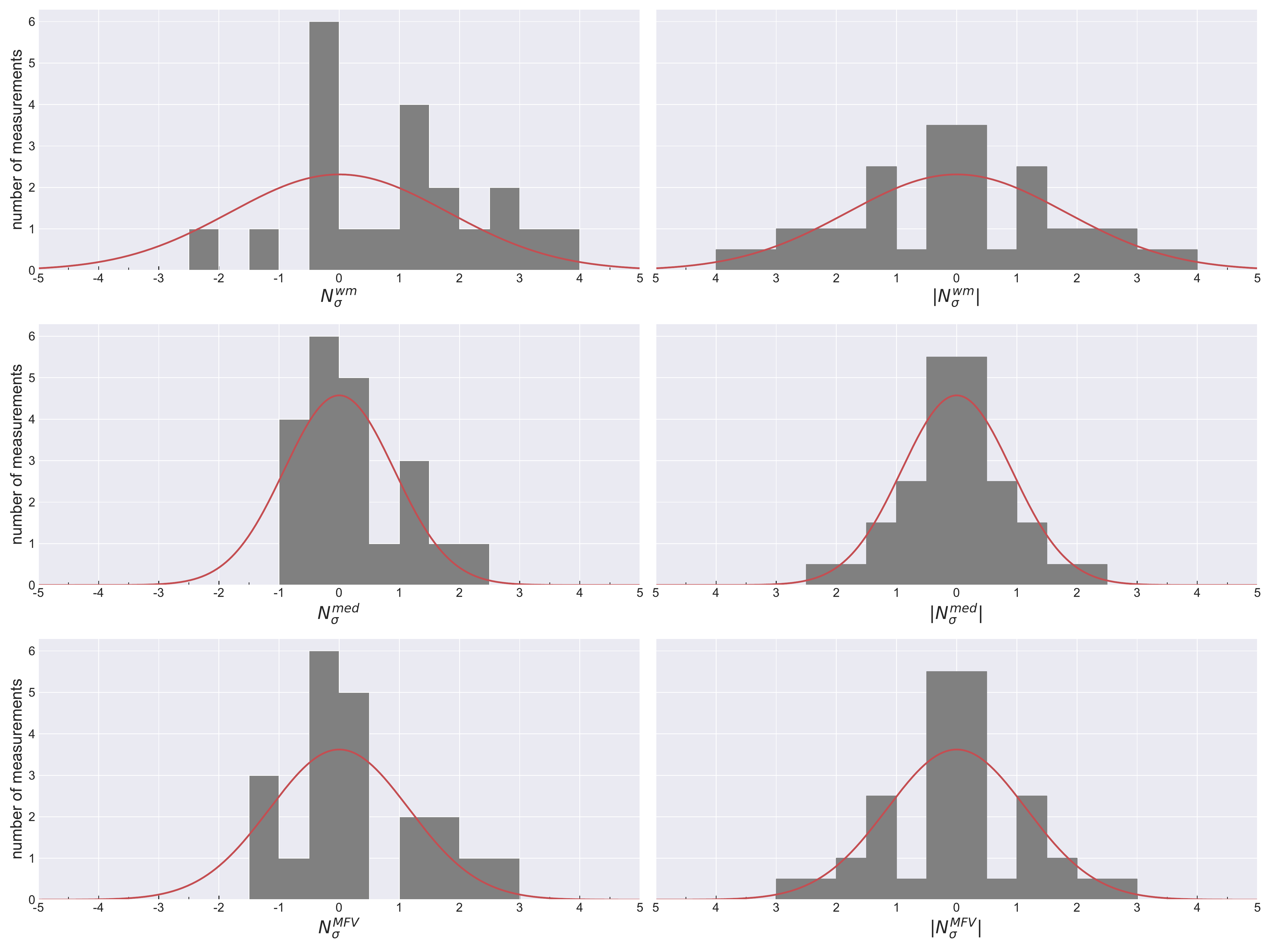}}
\end{center}
\vskip 0mm
\centerline{\footnotesize \begin{tabular}{p{10cm}}\bf Fig.\,6. \rm
		Histograms of the number of standard deviations in half bins away from the weighted mean, median, and MFV listed in the top, middle, and bottom rows.  The left (right) column illustrates the signed (absolute) deviation, where the smooth curves in panels represent the best-fit Gaussian. The  $N_\sigma$ of positive and negative cases indicate greater and less than the weighted mean, median, and MFV. [A colour version of this figure is available in the online version.]
\end{tabular}}
\vskip 0.5\baselineskip
Applying weighted mean statistics, the neutron lifetime measurements produce a central estimate of $\tau_n = 878.69\pm0.25$ s, while the PDG average is  $\tau_n = 878.4\pm 0.5$ s including scale factor of 1.8\cite{PDG22}.  
We also obtain $\chi^2=3.64$ and the number of standard deviations is $N=5.75$. 
Based on the median statistics, we have obtained a central estimate of $\tau_n = 881.5^{+5.5}_{-3}$ s with a $1\sigma$ range of $\left[878.5, 887.0\right]$. The MFV estimate is given by $\tau_n = 881.16^{+2.25}_{-2.35} $ s with uncertainty corresponding to 68.27\% confidence intervals.

\begin{center}
	\centerline{\includegraphics[width=1\textwidth]{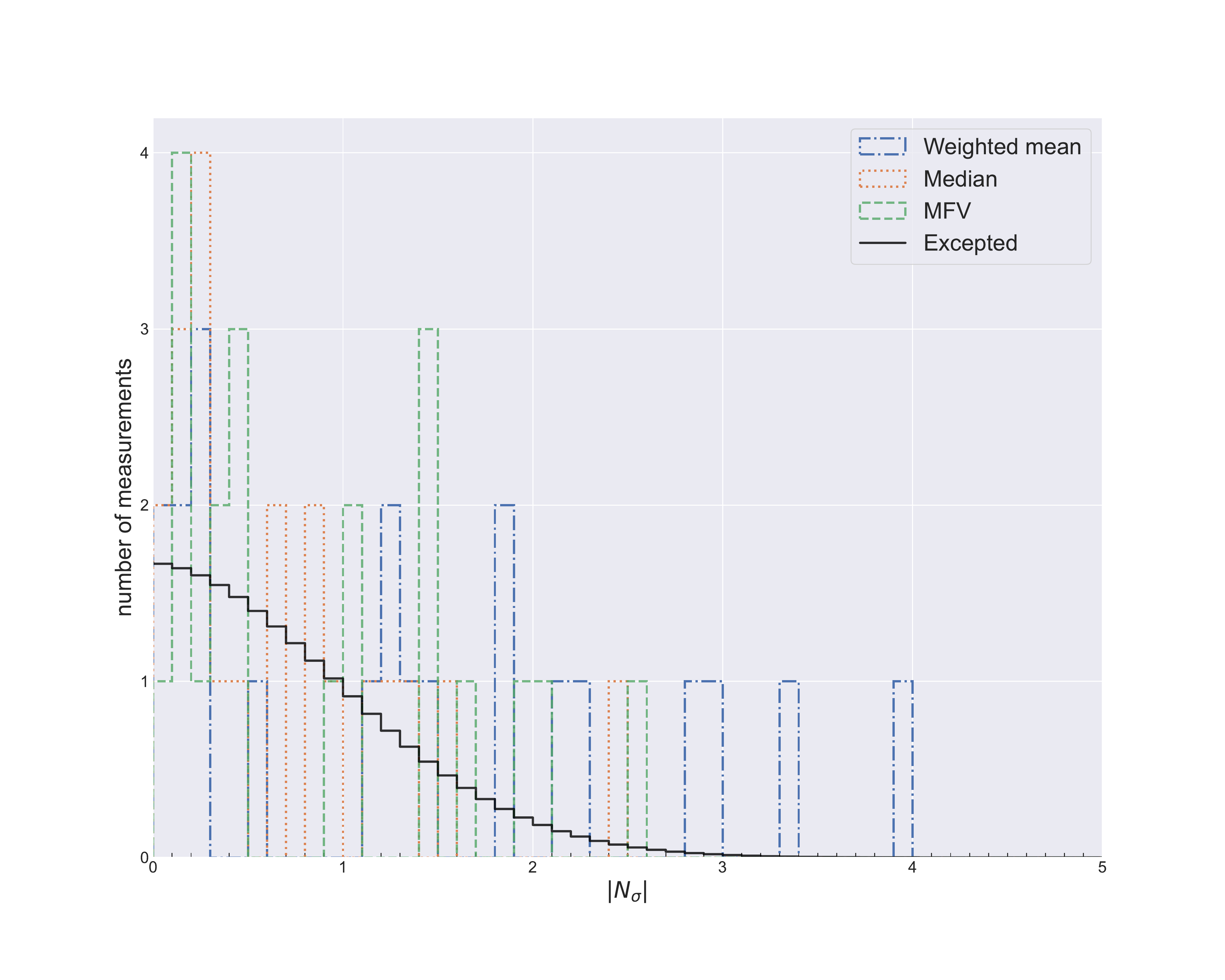}}
\end{center}
\vskip 0mm
\centerline{\footnotesize \begin{tabular}{p{12cm}}\bf Fig.\,7. \rm
		Histogram of the error distribution in $\mid N_\sigma \mid = 0.1$. The solid black line indicates the expected Gaussian probabilities for all data and the dash-dotted, dash, and dot lines denote the numbers of $\mid N_\sigma \mid$ values for the weighted mean, MFV, and median, respectively. [A colour version of this figure is available in the online version.]
\end{tabular}}

We now apply these statistical methods to plot the error distributions of neutron lifetime measurements illustrated as a function of $N_\sigma$\cite{Crandall2015ApJ}, Equation (11)-(13), in Fig. 6, which demonstrates the $N_\sigma$ and symmetrical $\mid N_\sigma\mid $ histograms using the weighted mean, median, and MFV. The histogram of error distributions of the measurements is shown in Fig. 7 with $\mid N_\sigma \mid =0.1$ bin size for a more specific viewpoint. As can be seen from these figures, the weighted mean case is wider than the expected Gaussian, which should generate a single value with  $\mid  N_\sigma \mid \ge 2$  and none value with $\mid  N_\sigma \mid \ge 3$. But there are 6 values with $\mid  N_\sigma \mid \ge 2$, 2 with $\mid  N_\sigma \mid \ge 3$, and none with $\mid  N_\sigma \mid \ge 4$.  Remarkably,   $68.3\%$ of the $N_{\sigma_{wm}}$ error distribution lies within $-1.24\le  N_\sigma \le 1.87$ while 95.4\% falls within $-2.22\le  N_\sigma \le 3.99$ .  The observed  $N_{\sigma_{wm}}$ error distribution has constraints of $\mid  N_\sigma \mid \le 1.87$ and $\mid  N_\sigma \mid \le 3.99$ respectively, and $38.1\%$ and  $71.4\%$ of the values lie within $\mid  N_\sigma \mid \le 1$ and $\mid  N_\sigma \mid \le 2$, respectively. 
For the median case, the distribution has a narrower tail than the expected Gaussian distribution, with $1$ value of $\mid  N_\sigma \mid \ge 2$ and none with $\mid  N_\sigma \mid \ge 3$. For signed $N_\sigma$, $68.3\%$ of the data lie within $-0.69\le  N_\sigma \le 1.29$ , while 95.4\% fall within $-0.88\le  N_\sigma \le 2.50$.  
The absolute $\mid N_\sigma \mid $ error distribution has constraints of $\mid  N_\sigma \mid \le 0.88$ and $\mid  N_\sigma \mid \le 2.50$, respectively. 
Moreover, $76.2\%$ and $95.2\%$ of the values lie within  $\mid N_\sigma \mid \le 1$  and  $\mid  N_\sigma \mid \le 2$ , respectively.  

On the other hand, for the MFV case, we gain a central estimate of $\tau_n=881.16$ s, see Fig. 8, and also find a non-Gaussian error distribution  with $2$ values of $\mid  N_\sigma \mid \ge 2$ and none with $\mid N_\sigma \mid \ge 3$.   
$68.3\%$ of the data falls within $-0.97\le N_\sigma \le 1.66$ , while 95.4\% lie within $-1.47\le  N_\sigma \le 2.60$.  
The $\mid  N_\sigma \mid $ error distribution has constraints of $\mid N_\sigma \mid \le 1.43$ and $\mid  N_\sigma \mid \le 2.60$, respectively, and $57.1\%$ and $90.5\%$ of the values fall within  $\mid  N_\sigma \mid \le 1$  and  $\mid  N_\sigma \mid \le 2$ , respectively. These data highlight that the error distribution for the MFV case is broader than that of the median case and narrower than that of the weighted mean case. The incrementally narrowing distributions of the weighted mean, MFV, and median might be related to unaccounted-for systematic uncertainties or correlations among measurements of neutron lifetime\cite{Crandall2015ApJ,de_Grijs2014AJa, Cowan2019}. Obviously, these detailed statistical descriptions represent that the error distributions of the weighted mean, median, and MFV are non-Gaussian, and the MFV technique is appropriate for a robust analysis of neutron lifetime measurements. In Fig. 8 we show all neutron lifetime measurements of bottle, beam, and space-based methods in the top panel, while the calculated residuals  $\bigtriangleup \tau / \tau_{MFV}$ from the MFV are demonstrated in the bottom panel. This provides a strong evidence for the non-Normality of error distributions, further confirming the effectiveness and rationality of MFV statistics. The data of neutron lifetime and codes to reproduce above results are available in the following repository: https://gitee.com/zhangphysics/neutron-lifetime.

\begin{center}
	\centerline{\includegraphics[width=1\textwidth]{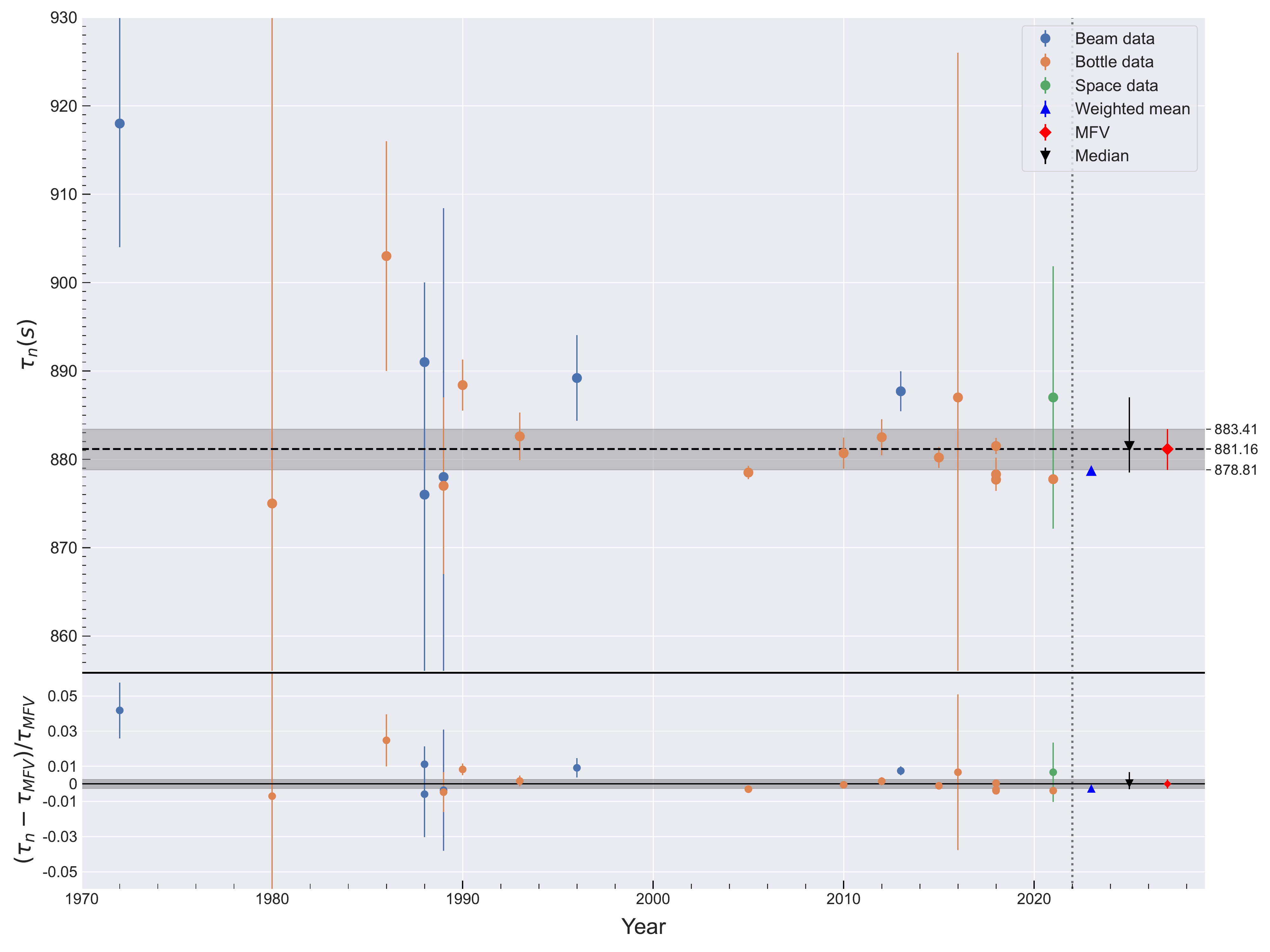}}
\end{center}
\vskip 0mm
\centerline{\footnotesize \begin{tabular}{p{12cm}}\bf Fig.\,8. \rm
		(Top) Neutron lifetime measurements as a function of publication date. The points are the measured data using bottle, beam, and space-based methods, while the triangle, diamond, and inverted triangle depict the weighted mean, MFV, and median, respectively.  (Bottom) We show the residuals of fit or data with respect to the MFV as a function of publication date, i.e. $\bigtriangleup \tau / \tau_{MFV}$. Data set key: PDG 2022 edition\cite{PDG22}.[A colour version of this figure is available in the online version.]
\end{tabular}}

\section{Conclusion}\label{sec5}

In summary, the problem of the long-standing tension in neutron lifetime measurements is one of the important challenges of astrophysics and particle physics. In this paper, from the perspective of robust statistics of the observed data, we have applied the MFV statistics technique to explore a detailed statistical analysis for the updated PDG dataset of neutron lifetime measurements that extend from beam to bottle methods. The MFV estimate is given by $\tau_n=881.16^{+2.25}_{-2.35}$ s with uncertainty corresponding to 68.27\% confidence interval, while the result of median statistics is $\tau_n=881.5^{+5.5}_{-3}$ s. 
Additionally, we have investigated the error distributions of neutron lifetime measurements compiled by PDG.
Due to the non-Gaussian error distributions, the weighted mean technique is more inappropriate than the median and MFV statistics. Similar to median statistics, the MFV statistics error distributions are clearly different from Gaussian, which implies that it might be caused by unaccounted-for systematic uncertainties or correlations among all data. As can be seen, the MFV technique is a powerful tool and appropriate for analysis of non-Gaussian distributions. Using the MFV statistics, we can gain insight into the details of how error distributions construct. Moreover, the consistent results demonstrated the usage and robustness of MFV statistics, which will inspire further uncertainty research into the application of the MFV method in some similar situations. 

\backmatter

\bmhead{Acknowledgments}
We greatly appreciate the anonymous referee for a careful reading and valuable comments that improve the paper. We are grateful to Z.-W Han, B. Zhang and E. Feigelson for valuable discussions. This work was supported by the National Natural Science Foundation of China (Grant No. 11547041, 11403007, 11701135, 11673007), the Natural Science Foundation of Hebei Province (A2017403025, A2021403002).

\bibliography{newmybib}


\begin{thebibliography}{72}
\ifx \bisbn   \undefined \def \bisbn  #1{ISBN #1}\fi
\ifx \binits  \undefined \def \binits#1{#1}\fi
\ifx \bauthor  \undefined \def \bauthor#1{#1}\fi
\ifx \batitle  \undefined \def \batitle#1{#1}\fi
\ifx \bjtitle  \undefined \def \bjtitle#1{#1}\fi
\ifx \bvolume  \undefined \def \bvolume#1{\textbf{#1}}\fi
\ifx \byear  \undefined \def \byear#1{#1}\fi
\ifx \bissue  \undefined \def \bissue#1{#1}\fi
\ifx \bfpage  \undefined \def \bfpage#1{#1}\fi
\ifx \blpage  \undefined \def \blpage #1{#1}\fi
\ifx \burl  \undefined \def \burl#1{\textsf{#1}}\fi
\ifx \doiurl  \undefined \def \doiurl#1{\url{https://doi.org/#1}}\fi
\ifx \betal  \undefined \def \betal{\textit{et al.}}\fi
\ifx \binstitute  \undefined \def \binstitute#1{#1}\fi
\ifx \binstitutionaled  \undefined \def \binstitutionaled#1{#1}\fi
\ifx \bctitle  \undefined \def \bctitle#1{#1}\fi
\ifx \beditor  \undefined \def \beditor#1{#1}\fi
\ifx \bpublisher  \undefined \def \bpublisher#1{#1}\fi
\ifx \bbtitle  \undefined \def \bbtitle#1{#1}\fi
\ifx \bedition  \undefined \def \bedition#1{#1}\fi
\ifx \bseriesno  \undefined \def \bseriesno#1{#1}\fi
\ifx \blocation  \undefined \def \blocation#1{#1}\fi
\ifx \bsertitle  \undefined \def \bsertitle#1{#1}\fi
\ifx \bsnm \undefined \def \bsnm#1{#1}\fi
\ifx \bsuffix \undefined \def \bsuffix#1{#1}\fi
\ifx \bparticle \undefined \def \bparticle#1{#1}\fi
\ifx \barticle \undefined \def \barticle#1{#1}\fi
\bibcommenthead
\ifx \bconfdate \undefined \def \bconfdate #1{#1}\fi
\ifx \botherref \undefined \def \botherref #1{#1}\fi
\ifx \url \undefined \def \url#1{\textsf{#1}}\fi
\ifx \bchapter \undefined \def \bchapter#1{#1}\fi
\ifx \bbook \undefined \def \bbook#1{#1}\fi
\ifx \bcomment \undefined \def \bcomment#1{#1}\fi
\ifx \oauthor \undefined \def \oauthor#1{#1}\fi
\ifx \citeauthoryear \undefined \def \citeauthoryear#1{#1}\fi
\ifx \endbibitem  \undefined \def \endbibitem {}\fi
\ifx \bconflocation  \undefined \def \bconflocation#1{#1}\fi
\ifx \arxivurl  \undefined \def \arxivurl#1{\textsf{#1}}\fi
\csname PreBibitemsHook\endcsname

\bibitem{Wietfeldt2011}
\begin{barticle}
\bauthor{\bsnm{{Wietfeldt}}, \binits{F.E.}},
\bauthor{\bsnm{{Greene}}, \binits{G.L.}}:
\batitle{{Colloquium: The neutron lifetime}}.
\bjtitle{Rev. Mod. Phys.}
\bvolume{83}(\bissue{4}),
\bfpage{1173}--\blpage{1192}
(\byear{2011}).
\doiurl{10.1103/RevModPhys.83.1173}
\end{barticle}
\endbibitem

\bibitem{Wietfeldt2018}
\begin{barticle}
\bauthor{\bsnm{{Wietfeldt}}, \binits{F.}}:
\batitle{{Measurements of the Neutron Lifetime}}.
\bjtitle{Atoms}
\bvolume{6}(\bissue{4}),
\bfpage{70}
(\byear{2018}).
\doiurl{10.3390/atoms6040070}
\end{barticle}
\endbibitem

\bibitem{Yue2013}
\begin{barticle}
\bauthor{\bsnm{{Yue}}, \binits{A.T.}},
\bauthor{\bsnm{{Dewey}}, \binits{M.S.}},
\bauthor{\bsnm{{Gilliam}}, \binits{D.M.}},
\bauthor{\bsnm{{Greene}}, \binits{G.L.}},
\bauthor{\bsnm{{Laptev}}, \binits{A.B.}},
\bauthor{\bsnm{{Nico}}, \binits{J.S.}},
\bauthor{\bsnm{{Snow}}, \binits{W.M.}},
\bauthor{\bsnm{{Wietfeldt}}, \binits{F.E.}}:
\batitle{{Improved Determination of the Neutron Lifetime}}.
\bjtitle{\prl}
\bvolume{111}(\bissue{22}),
\bfpage{222501}
(\byear{2013})
{\href{https://arxiv.org/abs/1309.2623}{{arXiv:1309.2623}}}
{[nucl-ex]}.
\doiurl{10.1103/PhysRevLett.111.222501}
\end{barticle}
\endbibitem

\bibitem{Huffman2000}
\begin{barticle}
\bauthor{\bsnm{{Huffman}}, \binits{P.R.}},
\bauthor{\bsnm{{Brome}}, \binits{C.R.}},
\bauthor{\bsnm{{Butterworth}}, \binits{J.S.}},
\bauthor{\bsnm{{Coakley}}, \binits{K.J.}},
\bauthor{\bsnm{{Dewey}}, \binits{M.S.}},
\bauthor{\bsnm{{Dzhosyuk}}, \binits{S.N.}},
\bauthor{\bsnm{{Golub}}, \binits{R.}},
\bauthor{\bsnm{{Greene}}, \binits{G.L.}},
\bauthor{\bsnm{{Habicht}}, \binits{K.}},
\bauthor{\bsnm{{Lamoreaux}}, \binits{S.K.}},
\bauthor{\bsnm{{Mattoni}}, \binits{C.E.H.}},
\bauthor{\bsnm{{McKinsey}}, \binits{D.N.}},
\bauthor{\bsnm{{Wietfeldt}}, \binits{F.E.}},
\bauthor{\bsnm{{Doyle}}, \binits{J.M.}}:
\batitle{{Magnetic trapping of neutrons}}.
\bjtitle{\nat}
\bvolume{403}(\bissue{6765}),
\bfpage{62}--\blpage{64}
(\byear{2000})
{\href{https://arxiv.org/abs/nucl-ex/0001003}{{arXiv:nucl-ex/0001003}}}
{[nucl-ex]}.
\doiurl{10.1038/47444}
\end{barticle}
\endbibitem

\bibitem{Byrne2019}
\begin{barticle}
\bauthor{\bsnm{{Byrne}}, \binits{J.}},
\bauthor{\bsnm{{Worcester}}, \binits{D.L.}}:
\batitle{{The neutron lifetime anomaly and charge exchange collisions of
  trapped protons}}.
\bjtitle{J. Phys. G Nucl. Phys.}
\bvolume{46}(\bissue{8}),
\bfpage{085001}
(\byear{2019}).
\doiurl{10.1088/1361-6471/ab256b}
\end{barticle}
\endbibitem

\bibitem{Serebrov2018}
\begin{barticle}
\bauthor{\bsnm{{Serebrov}}, \binits{A.P.}},
\bauthor{\bsnm{{Kolomensky}}, \binits{E.A.}},
\bauthor{\bsnm{{Fomin}}, \binits{A.K.}},
\bauthor{\bsnm{{Krasnoshchekova}}, \binits{I.A.}},
\bauthor{\bsnm{{Vassiljev}}, \binits{A.V.}},
\bauthor{\bsnm{{Prudnikov}}, \binits{D.M.}},
\bauthor{\bsnm{{Shoka}}, \binits{I.V.}},
\bauthor{\bsnm{{Chechkin}}, \binits{A.V.}},
\bauthor{\bsnm{{Chaikovskiy}}, \binits{M.E.}},
\bauthor{\bsnm{{Varlamov}}, \binits{V.E.}}:
\batitle{{Neutron lifetime measurements with a large gravitational trap for
  ultracold neutrons}}.
\bjtitle{\prc}
\bvolume{97}(\bissue{5}),
\bfpage{055503}
(\byear{2018}).
\doiurl{10.1103/PhysRevC.97.055503}
\end{barticle}
\endbibitem

\bibitem{Pattie2018}
\begin{barticle}
\bauthor{\bsnm{{Pattie}}, \binits{R.W.}},
\bauthor{\bsnm{{Callahan}}, \binits{N.B.}},
\bauthor{\bsnm{{Cude-Woods}}, \binits{C.}},
\bauthor{\bsnm{{Adamek}}, \binits{E.R.}},
\bauthor{\bsnm{{Broussard}}, \binits{L.J.}},
\bauthor{\bsnm{{Clayton}}, \binits{S.M.}},
\bauthor{\bsnm{{Currie}}, \binits{S.A.}},
\bauthor{\bsnm{{Dees}}, \binits{E.B.}},
\bauthor{\bsnm{{Ding}}, \binits{X.}},
\bauthor{\bsnm{{Engel}}, \binits{E.M.}}:
\batitle{{Measurement of the neutron lifetime using a magneto-gravitational
  trap and in situ detection}}.
\bjtitle{Science}
\bvolume{360}(\bissue{6389}),
\bfpage{627}--\blpage{632}
(\byear{2018})
{\href{https://arxiv.org/abs/1707.01817}{{arXiv:1707.01817}}}
{[nucl-ex]}.
\doiurl{10.1126/science.aan8895}
\end{barticle}
\endbibitem

\bibitem{Gonzalez2021}
\begin{barticle}
\bauthor{\bsnm{{Gonzalez}}, \binits{F.M.}},
\bauthor{\bsnm{{Fries}}, \binits{E.M.}},
\bauthor{\bsnm{{Cude-Woods}}, \binits{C.}},
\bauthor{\bsnm{{Bailey}}, \binits{T.}},
\bauthor{\bsnm{{Blatnik}}, \binits{M.}},
\bauthor{\bsnm{{Broussard}}, \binits{L.J.}},
\bauthor{\bsnm{{Callahan}}, \binits{N.B.}},
\bauthor{\bsnm{{Choi}}, \binits{J.H.}},
\bauthor{\bsnm{{Clayton}}, \binits{S.M.}},
\bauthor{\bsnm{{Currie}}, \binits{S.A.}},
\bauthor{\bsnm{{Dawid}}, \binits{M.}},
\bauthor{\bsnm{{Dees}}, \binits{E.B.}},
\bauthor{\bsnm{{Filippone}}, \binits{B.W.}},
\bauthor{\bsnm{{Fox}}, \binits{W.}},
\bauthor{\bsnm{{Geltenbort}}, \binits{P.}},
\bauthor{\bsnm{{George}}, \binits{E.}},
\bauthor{\bsnm{{Hayen}}, \binits{L.}},
\bauthor{\bsnm{{Hickerson}}, \binits{K.P.}},
\bauthor{\bsnm{{Hoffbauer}}, \binits{M.A.}},
\bauthor{\bsnm{{Hoffman}}, \binits{K.}},
\bauthor{\bsnm{{Holley}}, \binits{A.T.}},
\bauthor{\bsnm{{Ito}}, \binits{T.M.}},
\bauthor{\bsnm{{Komives}}, \binits{A.}},
\bauthor{\bsnm{{Liu}}, \binits{C.-Y.}},
\bauthor{\bsnm{{Makela}}, \binits{M.}},
\bauthor{\bsnm{{Morris}}, \binits{C.L.}},
\bauthor{\bsnm{{Musedinovic}}, \binits{R.}},
\bauthor{\bsnm{{O'Shaughnessy}}, \binits{C.}},
\bauthor{\bsnm{{Pattie}}, \binits{R.W.}},
\bauthor{\bsnm{{Ramsey}}, \binits{J.}},
\bauthor{\bsnm{{Salvat}}, \binits{D.J.}},
\bauthor{\bsnm{{Saunders}}, \binits{A.}},
\bauthor{\bsnm{{Sharapov}}, \binits{E.I.}},
\bauthor{\bsnm{{Slutsky}}, \binits{S.}},
\bauthor{\bsnm{{Su}}, \binits{V.}},
\bauthor{\bsnm{{Sun}}, \binits{X.}},
\bauthor{\bsnm{{Swank}}, \binits{C.}},
\bauthor{\bsnm{{Tang}}, \binits{Z.}},
\bauthor{\bsnm{{Uhrich}}, \binits{W.}},
\bauthor{\bsnm{{Vanderwerp}}, \binits{J.}},
\bauthor{\bsnm{{Walstrom}}, \binits{P.}},
\bauthor{\bsnm{{Wang}}, \binits{Z.}},
\bauthor{\bsnm{{Wei}}, \binits{W.}},
\bauthor{\bsnm{{Young}}, \binits{A.R.}},
\bauthor{\bsnm{{UCN {\ensuremath{\tau}} Collaboration}}}:
\batitle{{Improved Neutron Lifetime Measurement with UCN {\ensuremath{\tau}}}}.
\bjtitle{\prl}
\bvolume{127}(\bissue{16}),
\bfpage{162501}
(\byear{2021})
{\href{https://arxiv.org/abs/2106.10375}{{arXiv:2106.10375}}}
{[nucl-ex]}.
\doiurl{10.1103/PhysRevLett.127.162501}
\end{barticle}
\endbibitem

\bibitem{Wilson2021}
\begin{barticle}
\bauthor{\bsnm{{Wilson}}, \binits{J.T.}},
\bauthor{\bsnm{{Lawrence}}, \binits{D.J.}},
\bauthor{\bsnm{{Peplowski}}, \binits{P.N.}},
\bauthor{\bsnm{{Eke}}, \binits{V.R.}},
\bauthor{\bsnm{{Kegerreis}}, \binits{J.A.}}:
\batitle{{Measurement of the free neutron lifetime using the neutron
  spectrometer on NASA's Lunar Prospector mission}}.
\bjtitle{\prc}
\bvolume{104}(\bissue{4}),
\bfpage{045501}
(\byear{2021})
{\href{https://arxiv.org/abs/2011.07061}{{arXiv:2011.07061}}}
{[nucl-ex]}.
\doiurl{10.1103/PhysRevC.104.045501}
\end{barticle}
\endbibitem

\bibitem{Fornal2018}
\begin{barticle}
\bauthor{\bsnm{{Fornal}}, \binits{B.}},
\bauthor{\bsnm{{Grinstein}}, \binits{B.}}:
\batitle{{Dark Matter Interpretation of the Neutron Decay Anomaly}}.
\bjtitle{\prl}
\bvolume{120}(\bissue{19}),
\bfpage{191801}
(\byear{2018})
{\href{https://arxiv.org/abs/1811.03086}{{arXiv:1811.03086}}}
{[hep-ph]}.
\doiurl{10.1103/PhysRevLett.120.191801}
\end{barticle}
\endbibitem

\bibitem{Gott2001}
\begin{barticle}
\bauthor{\bsnm{{Gott}}, \binits{I.} \bsuffix{J.~Richard}},
\bauthor{\bsnm{{Vogeley}}, \binits{M.S.}},
\bauthor{\bsnm{{Podariu}}, \binits{S.}},
\bauthor{\bsnm{{Ratra}}, \binits{B.}}:
\batitle{{Median Statistics, H$_{0}$, and the Accelerating Universe}}.
\bjtitle{\apj}
\bvolume{549}(\bissue{1}),
\bfpage{1}--\blpage{17}
(\byear{2001})
{\href{https://arxiv.org/abs/astro-ph/0006103}{{arXiv:astro-ph/0006103}}}
{[astro-ph]}.
\doiurl{10.1086/319055}
\end{barticle}
\endbibitem

\bibitem{Bailey2017}
\begin{barticle}
\bauthor{\bsnm{{Bailey}}, \binits{D.C.}}:
\batitle{{Not Normal: the uncertainties of scientific measurements}}.
\bjtitle{R. Soc. Open Sci.}
\bvolume{4}(\bissue{1}),
\bfpage{160600}
(\byear{2017})
{\href{https://arxiv.org/abs/1612.00778}{{arXiv:1612.00778}}}
{[stat.AP]}.
\doiurl{10.1098/rsos.160600}
\end{barticle}
\endbibitem

\bibitem{Chen2003a}
\begin{barticle}
\bauthor{\bsnm{{Chen}}, \binits{G.}},
\bauthor{\bsnm{{Ratra}}, \binits{B.}}:
\batitle{{Median Statistics and the Mass Density of the Universe}}.
\bjtitle{\pasp}
\bvolume{115},
\bfpage{1143}--\blpage{1149}
(\byear{2003})
{\href{https://arxiv.org/abs/astro-ph/0302002}{{astro-ph/0302002}}}.
\doiurl{10.1086/377112}
\end{barticle}
\endbibitem

\bibitem{Chen2011}
\begin{barticle}
\bauthor{\bsnm{{Chen}}, \binits{G.}},
\bauthor{\bsnm{{Ratra}}, \binits{B.}}:
\batitle{{Median Statistics and the Hubble Constant}}.
\bjtitle{\pasp}
\bvolume{123}(\bissue{907}),
\bfpage{1127}
(\byear{2011})
{\href{https://arxiv.org/abs/1105.5206}{{arXiv:1105.5206}}}
{[astro-ph.CO]}.
\doiurl{10.1086/662131}
\end{barticle}
\endbibitem

\bibitem{Crandall2015MPLA}
\begin{barticle}
\bauthor{\bsnm{{Crandall}}, \binits{S.}},
\bauthor{\bsnm{{Houston}}, \binits{S.}},
\bauthor{\bsnm{{Ratra}}, \binits{B.}}:
\batitle{{Non-Gaussian error distribution of 7Li abundance measurements}}.
\bjtitle{Mod. Phys. Lett. A}
\bvolume{30}(\bissue{25}),
\bfpage{1550123}
(\byear{2015})
{\href{https://arxiv.org/abs/1409.7332}{{arXiv:1409.7332}}}
{[astro-ph.CO]}.
\doiurl{10.1142/S0217732315501230}
\end{barticle}
\endbibitem

\bibitem{Zhang2017}
\begin{barticle}
\bauthor{\bsnm{{Zhang}}, \binits{J.}}:
\batitle{{Most frequent value statistics and distribution of $^{7}$Li abundance
  observations}}.
\bjtitle{\mnras}
\bvolume{468},
\bfpage{5014}--\blpage{5019}
(\byear{2017}).
\doiurl{10.1093/mnras/stx627}
\end{barticle}
\endbibitem

\bibitem{Crandall2015ApJ}
\begin{barticle}
\bauthor{\bsnm{{Crandall}}, \binits{S.}},
\bauthor{\bsnm{{Ratra}}, \binits{B.}}:
\batitle{{Non-Gaussian Error Distributions of LMC Distance Moduli
  Measurements}}.
\bjtitle{\apj}
\bvolume{815}(\bissue{2}),
\bfpage{87}
(\byear{2015})
{\href{https://arxiv.org/abs/1507.07940}{{arXiv:1507.07940}}}
{[astro-ph.CO]}.
\doiurl{10.1088/0004-637X/815/2/87}
\end{barticle}
\endbibitem

\bibitem{Penton2018}
\begin{barticle}
\bauthor{\bsnm{{Penton}}, \binits{J.}},
\bauthor{\bsnm{{Peyton}}, \binits{J.}},
\bauthor{\bsnm{{Zahoor}}, \binits{A.}},
\bauthor{\bsnm{{Ratra}}, \binits{B.}}:
\batitle{{Median Statistics Analysis of Deuterium Abundance Measurements and
  Spatial Curvature Constraints}}.
\bjtitle{\pasp}
\bvolume{130}(\bissue{993}),
\bfpage{114001}
(\byear{2018})
{\href{https://arxiv.org/abs/1808.01490}{{arXiv:1808.01490}}}
{[astro-ph.CO]}.
\doiurl{10.1088/1538-3873/aadf75}
\end{barticle}
\endbibitem

\bibitem{Camarillo2018PASP}
\begin{barticle}
\bauthor{\bsnm{{Camarillo}}, \binits{T.}},
\bauthor{\bsnm{{Mathur}}, \binits{V.}},
\bauthor{\bsnm{{Mitchell}}, \binits{T.}},
\bauthor{\bsnm{{Ratra}}, \binits{B.}}:
\batitle{{Median Statistics Estimate of the Distance to the Galactic Center}}.
\bjtitle{\pasp}
\bvolume{130}(\bissue{984}),
\bfpage{024101}
(\byear{2018})
{\href{https://arxiv.org/abs/1708.01310}{{arXiv:1708.01310}}}
{[astro-ph.GA]}.
\doiurl{10.1088/1538-3873/aa9b26}
\end{barticle}
\endbibitem

\bibitem{Camarillo2018Apss}
\begin{barticle}
\bauthor{\bsnm{{Camarillo}}, \binits{T.}},
\bauthor{\bsnm{{Dredger}}, \binits{P.}},
\bauthor{\bsnm{{Ratra}}, \binits{B.}}:
\batitle{{Median statistics estimate of the galactic rotational velocity}}.
\bjtitle{\apss}
\bvolume{363}(\bissue{12}),
\bfpage{268}
(\byear{2018})
{\href{https://arxiv.org/abs/1805.01917}{{arXiv:1805.01917}}}
{[astro-ph.GA]}.
\doiurl{10.1007/s10509-018-3486-8}
\end{barticle}
\endbibitem

\bibitem{Rajan2018}
\begin{barticle}
\bauthor{\bsnm{{Rajan}}, \binits{A.}},
\bauthor{\bsnm{{Desai}}, \binits{S.}}:
\batitle{{Non-Gaussian error distributions of galactic rotation speed
  measurements}}.
\bjtitle{Eur. Phys. J. Plus}
\bvolume{133}(\bissue{3}),
\bfpage{107}
(\byear{2018})
{\href{https://arxiv.org/abs/1710.06624}{{arXiv:1710.06624}}}
{[astro-ph.IM]}.
\doiurl{10.1140/epjp/i2018-11946-7}
\end{barticle}
\endbibitem

\bibitem{Bethapudi2017}
\begin{barticle}
\bauthor{\bsnm{{Bethapudi}}, \binits{S.}},
\bauthor{\bsnm{{Desai}}, \binits{S.}}:
\batitle{{Median statistics estimates of Hubble and Newton's constants}}.
\bjtitle{Eur. Phys. J. Plus}
\bvolume{132},
\bfpage{78}
(\byear{2017})
{\href{https://arxiv.org/abs/1701.01789}{{arXiv:1701.01789}}}.
\doiurl{10.1140/epjp/i2017-11390-3}
\end{barticle}
\endbibitem

\bibitem{Rajan2020}
\begin{barticle}
\bauthor{\bsnm{{Rajan}}, \binits{A.}},
\bauthor{\bsnm{{Desai}}, \binits{S.}}:
\batitle{{A meta-analysis of neutron lifetime measurements}}.
\bjtitle{Progress of Theoretical and Experimental Physics}
\bvolume{2020}(\bissue{1}),
\bfpage{013}--\blpage{01}
(\byear{2020})
{\href{https://arxiv.org/abs/1812.09671}{{arXiv:1812.09671}}}
{[hep-ph]}.
\doiurl{10.1093/ptep/ptz153}
\end{barticle}
\endbibitem

\bibitem{Tanabashi2018}
\begin{barticle}
\bauthor{\bsnm{{Tanabashi}}, \binits{M.}},
\bauthor{\bsnm{{Hagiwara}}, \binits{K.}},
\bauthor{\bsnm{{Hikasa}}, \binits{K.}},
\bauthor{\bsnm{{Nakamura}}, \binits{K.}},
\bauthor{\bsnm{{Sumino}}, \binits{Y.}},
\bauthor{\bsnm{{Takahashi}}, \binits{F.}},
\bauthor{\bsnm{{Tanaka}}, \binits{J.}},
\bauthor{\bsnm{{Agashe}}, \binits{K.}},
\bauthor{\bsnm{{Aielli}}, \binits{G.}},
\bauthor{\bsnm{{Amsler}}, \binits{C.}}:
\batitle{{Review of Particle Physics$^{*}$}}.
\bjtitle{\prd}
\bvolume{98}(\bissue{3}),
\bfpage{030001}
(\byear{2018}).
\doiurl{10.1103/PhysRevD.98.030001}
\end{barticle}
\endbibitem

\bibitem{Steiner1988}
\begin{barticle}
\bauthor{\bsnm{Steiner}, \binits{F.}}:
\batitle{Most frequent value procedures}.
\bjtitle{Geophys. Trans.}
\bvolume{34}(\bissue{2-3}),
\bfpage{139}--\blpage{260}
(\byear{1988})
\end{barticle}
\endbibitem

\bibitem{Steiner1991}
\begin{bbook}
\beditor{\bsnm{Steiner}, \binits{F.}} (ed.):
\bbtitle{The Most Frequent Value. In Troduction to Modern Conception
  Statistics.}
\bpublisher{Akademia Kiado, Budapest, Hungary}, \blocation{ }
(\byear{1991})
\end{bbook}
\endbibitem

\bibitem{Steiner1997}
\begin{bbook}
\beditor{\bsnm{Steiner}, \binits{F.}} (ed.):
\bbtitle{Optimum Methods in Statistics}.
\bpublisher{Akademia Kiado, Budapest, Hungary}, \blocation{ }
(\byear{1997})
\end{bbook}
\endbibitem

\bibitem{Steiner2001}
\begin{barticle}
\bauthor{\bsnm{Steiner}, \binits{F.}},
\bauthor{\bsnm{Hajagos}, \binits{B.}}
\bjtitle{Acta Geod Geoph Hung}
\bvolume{36},
\bfpage{327}
(\byear{2001})
\end{barticle}
\endbibitem

\bibitem{Kemp2006}
\begin{bbook}
\bauthor{\bsnm{Kemp}, \binits{A.W.}}:
\bbtitle{Steiner's Most Frequent Value. Encyclopedia of Statistical Sciences.
  12.}
\bpublisher{John Wiley \& Sons, Inc.}, \blocation{ }
(\byear{2006})
\end{bbook}
\endbibitem

\bibitem{Szucs2006}
\begin{barticle}
\bauthor{\bsnm{Szucs}, \binits{P.}},
\bauthor{\bsnm{Civan}, \binits{F.}},
\bauthor{\bsnm{Virag}, \binits{M.}}:
\batitle{Applicability of the most frequent value method in groundwater
  modeling}.
\bjtitle{Hydrogeol. J.}
\bvolume{14}(\bissue{1}),
\bfpage{31}--\blpage{43}
(\byear{2006})
\end{barticle}
\endbibitem

\bibitem{Szegedi2013}
\begin{barticle}
\bauthor{\bsnm{Szegedi}, \binits{H.}}
\bjtitle{Geosci. Eng.}
\bvolume{2}(\bissue{4}),
\bfpage{102}--\blpage{115}
(\byear{2013})
\end{barticle}
\endbibitem

\bibitem{Szegedi2014}
\begin{barticle}
\bauthor{\bsnm{Szegedi}, \binits{H.}},
\bauthor{\bsnm{Dobroka}, \binits{M.}}:
\batitle{On the use of steiner’s weights in inversion-based fourier
  transformation: robustification of a previously published algorithm}.
\bjtitle{Acta Geod Geophys.}
\bvolume{49},
\bfpage{95}--\blpage{104}
(\byear{2014})
\end{barticle}
\endbibitem

\bibitem{Szabo2018}
\begin{barticle}
\bauthor{\bsnm{{Szab{\'o}}}, \binits{N.P.}},
\bauthor{\bsnm{{Balogh}}, \binits{G.P.}},
\bauthor{\bsnm{{Stickel}}, \binits{J.}}:
\batitle{{Most frequent value-based factor analysis of direct-push logging
  data}}.
\bjtitle{Geophys. Prospect.}
\bvolume{66}(\bissue{3}),
\bfpage{530}--\blpage{548}
(\byear{2018}).
\doiurl{10.1111/1365-2478.12573}
\end{barticle}
\endbibitem

\bibitem{Zhang2018}
\begin{barticle}
\bauthor{\bsnm{{Zhang}}, \binits{J.}}:
\batitle{{Most Frequent Value Statistics and the Hubble Constant}}.
\bjtitle{\pasp}
\bvolume{130}(\bissue{990}),
\bfpage{084502}
(\byear{2018}).
\doiurl{10.1088/1538-3873/aac767}
\end{barticle}
\endbibitem

\bibitem{Ezhov2018}
\begin{barticle}
\bauthor{\bsnm{{Ezhov}}, \binits{V.F.}},
\bauthor{\bsnm{{Andreev}}, \binits{A.Z.}},
\bauthor{\bsnm{{Ban}}, \binits{G.}},
\bauthor{\bsnm{{Bazarov}}, \binits{B.A.}},
\bauthor{\bsnm{{Geltenbort}}, \binits{P.}},
\bauthor{\bsnm{{Glushkov}}, \binits{A.G.}},
\bauthor{\bsnm{{Knyazkov}}, \binits{V.A.}},
\bauthor{\bsnm{{Kovrizhnykh}}, \binits{N.A.}},
\bauthor{\bsnm{{Krygin}}, \binits{G.B.}},
\bauthor{\bsnm{{Naviliat-Cuncic}}, \binits{O.}}:
\batitle{{Measurement of the Neutron Lifetime with Ultracold Neutrons Stored in
  a Magneto-Gravitational Trap}}.
\bjtitle{Sov. J. Exp. Theor. Phys. Lett.}
\bvolume{107}(\bissue{11}),
\bfpage{671}--\blpage{675}
(\byear{2018}).
\doiurl{10.1134/S0021364018110024}
\end{barticle}
\endbibitem

\bibitem{Leung2016}
\begin{barticle}
\bauthor{\bsnm{{Leung}}, \binits{K.K.H.}},
\bauthor{\bsnm{{Geltenbort}}, \binits{P.}},
\bauthor{\bsnm{{Ivanov}}, \binits{S.}},
\bauthor{\bsnm{{Rosenau}}, \binits{F.}},
\bauthor{\bsnm{{Zimmer}}, \binits{O.}}:
\batitle{{Neutron lifetime measurements and effective spectral cleaning with an
  ultracold neutron trap using a vertical Halbach octupole permanent magnet
  array}}.
\bjtitle{\prc}
\bvolume{94}(\bissue{4}),
\bfpage{045502}
(\byear{2016})
{\href{https://arxiv.org/abs/1606.00929}{{arXiv:1606.00929}}}
{[nucl-ex]}.
\doiurl{10.1103/PhysRevC.94.045502}
\end{barticle}
\endbibitem

\bibitem{Arzumanov2015}
\begin{barticle}
\bauthor{\bsnm{{Arzumanov}}, \binits{S.}},
\bauthor{\bsnm{{Bondarenko}}, \binits{L.}},
\bauthor{\bsnm{{Chernyavsky}}, \binits{S.}},
\bauthor{\bsnm{{Geltenbort}}, \binits{P.}},
\bauthor{\bsnm{{Morozov}}, \binits{V.}},
\bauthor{\bsnm{{Nesvizhevsky}}, \binits{V.V.}},
\bauthor{\bsnm{{Panin}}, \binits{Y.}},
\bauthor{\bsnm{{Strepetov}}, \binits{A.}}:
\batitle{{A measurement of the neutron lifetime using the method of storage of
  ultracold neutrons and detection of inelastically up-scattered neutrons}}.
\bjtitle{Phys. Lett. B}
\bvolume{745},
\bfpage{79}--\blpage{89}
(\byear{2015}).
\doiurl{10.1016/j.physletb.2015.04.021}
\end{barticle}
\endbibitem

\bibitem{Steyerl2012}
\begin{barticle}
\bauthor{\bsnm{{Steyerl}}, \binits{A.}},
\bauthor{\bsnm{{Pendlebury}}, \binits{J.M.}},
\bauthor{\bsnm{{Kaufman}}, \binits{C.}},
\bauthor{\bsnm{{Malik}}, \binits{S.S.}},
\bauthor{\bsnm{{Desai}}, \binits{A.M.}}:
\batitle{{Quasielastic scattering in the interaction of ultracold neutrons with
  a liquid wall and application in a reanalysis of the Mambo I neutron-lifetime
  experiment}}.
\bjtitle{\prc}
\bvolume{85}(\bissue{6}),
\bfpage{065503}
(\byear{2012}).
\doiurl{10.1103/PhysRevC.85.065503}
\end{barticle}
\endbibitem

\bibitem{Pichlmaier2010}
\begin{barticle}
\bauthor{\bsnm{{Pichlmaier}}, \binits{A.}},
\bauthor{\bsnm{{Varlamov}}, \binits{V.}},
\bauthor{\bsnm{{Schreckenbach}}, \binits{K.}},
\bauthor{\bsnm{{Geltenbort}}, \binits{P.}}:
\batitle{{Neutron lifetime measurement with the UCN trap-in-trap MAMBO II}}.
\bjtitle{Phys. Lett. B}
\bvolume{693}(\bissue{3}),
\bfpage{221}--\blpage{226}
(\byear{2010}).
\doiurl{10.1016/j.physletb.2010.08.032}
\end{barticle}
\endbibitem

\bibitem{PDG22}
\begin{botherref}
\oauthor{\bsnm{Workman}, \binits{R.L.}},
\oauthor{\bsnm{Burkert}, \binits{V.D.}},
\oauthor{\bsnm{Crede}, \binits{V.}},
\oauthor{\bsnm{Klempt}, \binits{E.}},
\oauthor{\bsnm{Thoma}, \binits{U.}},
\oauthor{\bsnm{Tiator}, \binits{L.}},
\oauthor{\bsnm{Agashe}, \binits{K.}},
\oauthor{\bsnm{Aielli}, \binits{G.}},
\oauthor{\bsnm{Allanach}, \binits{B.C.}},
\oauthor{\bsnm{Amsler}, \binits{C.}},
\oauthor{\bsnm{Antonelli}, \binits{M.}},
\oauthor{\bsnm{Aschenauer}, \binits{E.C.}},
\oauthor{\bsnm{Asner}, \binits{D.M.}},
\oauthor{\bsnm{Baer}, \binits{H.}},
\oauthor{\bsnm{Banerjee}, \binits{S.}},
\oauthor{\bsnm{Barnett}, \binits{R.M.}},
\oauthor{\bsnm{Baudis}, \binits{L.}},
\oauthor{\bsnm{Bauer}, \binits{C.W.}},
\oauthor{\bsnm{Beatty}, \binits{J.J.}},
\oauthor{\bsnm{Belousov}, \binits{V.I.}},
\oauthor{\bsnm{Beringer}, \binits{J.}},
\oauthor{\bsnm{Bettini}, \binits{A.}},
\oauthor{\bsnm{Biebel}, \binits{O.}},
\oauthor{\bsnm{Black}, \binits{K.M.}},
\oauthor{\bsnm{Blucher}, \binits{E.}},
\oauthor{\bsnm{Bonventre}, \binits{R.}},
\oauthor{\bsnm{Bryzgalov}, \binits{V.V.}},
\oauthor{\bsnm{Buchmuller}, \binits{O.}},
\oauthor{\bsnm{Bychkov}, \binits{M.A.}},
\oauthor{\bsnm{Cahn}, \binits{R.N.}},
\oauthor{\bsnm{Carena}, \binits{M.}},
\oauthor{\bsnm{Ceccucci}, \binits{A.}},
\oauthor{\bsnm{Cerri}, \binits{A.}},
\oauthor{\bsnm{Chivukula}, \binits{R.S.}},
\oauthor{\bsnm{Cowan}, \binits{G.}},
\oauthor{\bsnm{Cranmer}, \binits{K.}},
\oauthor{\bsnm{Cremonesi}, \binits{O.}},
\oauthor{\bsnm{D'Ambrosio}, \binits{G.}},
\oauthor{\bsnm{Damour}, \binits{T.}},
\oauthor{\bparticle{de} \bsnm{Florian}, \binits{D.}},
\oauthor{\bparticle{de} \bsnm{Gouvêa}, \binits{A.}},
\oauthor{\bsnm{DeGrand}, \binits{T.}},
\oauthor{\bparticle{de} \bsnm{Jong}, \binits{P.}},
\oauthor{\bsnm{Demers}, \binits{S.}},
\oauthor{\bsnm{Dobrescu}, \binits{B.A.}},
\oauthor{\bsnm{D'Onofrio}, \binits{M.}},
\oauthor{\bsnm{Doser}, \binits{M.}},
\oauthor{\bsnm{Dreiner}, \binits{H.K.}},
\oauthor{\bsnm{Eerola}, \binits{P.}},
\oauthor{\bsnm{Egede}, \binits{U.}},
\oauthor{\bsnm{Eidelman}, \binits{S.}},
\oauthor{\bsnm{El-Khadra}, \binits{A.X.}},
\oauthor{\bsnm{Ellis}, \binits{J.}},
\oauthor{\bsnm{Eno}, \binits{S.C.}},
\oauthor{\bsnm{Erler}, \binits{J.}},
\oauthor{\bsnm{Ezhela}, \binits{V.V.}},
\oauthor{\bsnm{Fetscher}, \binits{W.}},
\oauthor{\bsnm{Fields}, \binits{B.D.}},
\oauthor{\bsnm{Freitas}, \binits{A.}},
\oauthor{\bsnm{Gallagher}, \binits{H.}},
\oauthor{\bsnm{Gershtein}, \binits{Y.}},
\oauthor{\bsnm{Gherghetta}, \binits{T.}},
\oauthor{\bsnm{Gonzalez-Garcia}, \binits{M.C.}},
\oauthor{\bsnm{Goodman}, \binits{M.}},
\oauthor{\bsnm{Grab}, \binits{C.}},
\oauthor{\bsnm{Gritsan}, \binits{A.V.}},
\oauthor{\bsnm{Grojean}, \binits{C.}},
\oauthor{\bsnm{Groom}, \binits{D.E.}},
\oauthor{\bsnm{Grünewald}, \binits{M.}},
\oauthor{\bsnm{Gurtu}, \binits{A.}},
\oauthor{\bsnm{Gutsche}, \binits{T.}},
\oauthor{\bsnm{Haber}, \binits{H.E.}},
\oauthor{\bsnm{Hamel}, \binits{M.}},
\oauthor{\bsnm{Hanhart}, \binits{C.}},
\oauthor{\bsnm{Hashimoto}, \binits{S.}},
\oauthor{\bsnm{Hayato}, \binits{Y.}},
\oauthor{\bsnm{Hebecker}, \binits{A.}},
\oauthor{\bsnm{Heinemeyer}, \binits{S.}},
\oauthor{\bsnm{Hernández-Rey}, \binits{J.J.}},
\oauthor{\bsnm{Hikasa}, \binits{K.}},
\oauthor{\bsnm{Hisano}, \binits{J.}},
\oauthor{\bsnm{Höcker}, \binits{A.}},
\oauthor{\bsnm{Holder}, \binits{J.}},
\oauthor{\bsnm{Hsu}, \binits{L.}},
\oauthor{\bsnm{Huston}, \binits{J.}},
\oauthor{\bsnm{Hyodo}, \binits{T.}},
\oauthor{\bsnm{Ianni}, \binits{A.}},
\oauthor{\bsnm{Kado}, \binits{M.}},
\oauthor{\bsnm{Karliner}, \binits{M.}},
\oauthor{\bsnm{Katz}, \binits{U.F.}},
\oauthor{\bsnm{Kenzie}, \binits{M.}},
\oauthor{\bsnm{Khoze}, \binits{V.A.}},
\oauthor{\bsnm{Klein}, \binits{S.R.}},
\oauthor{\bsnm{Krauss}, \binits{F.}},
\oauthor{\bsnm{Kreps}, \binits{M.}},
\oauthor{\bsnm{Križan}, \binits{P.}},
\oauthor{\bsnm{Krusche}, \binits{B.}},
\oauthor{\bsnm{Kwon}, \binits{Y.}},
\oauthor{\bsnm{Lahav}, \binits{O.}},
\oauthor{\bsnm{Laiho}, \binits{J.}},
\oauthor{\bsnm{Lellouch}, \binits{L.P.}},
\oauthor{\bsnm{Lesgourgues}, \binits{J.}},
\oauthor{\bsnm{Liddle}, \binits{A.R.}},
\oauthor{\bsnm{Ligeti}, \binits{Z.}},
\oauthor{\bsnm{Lin}, \binits{C.-J.}},
\oauthor{\bsnm{Lippmann}, \binits{C.}},
\oauthor{\bsnm{Liss}, \binits{T.M.}},
\oauthor{\bsnm{Littenberg}, \binits{L.}},
\oauthor{\bsnm{Lourenço}, \binits{C.}},
\oauthor{\bsnm{Lugovsky}, \binits{K.S.}},
\oauthor{\bsnm{Lugovsky}, \binits{S.B.}},
\oauthor{\bsnm{Lusiani}, \binits{A.}},
\oauthor{\bsnm{Makida}, \binits{Y.}},
\oauthor{\bsnm{Maltoni}, \binits{F.}},
\oauthor{\bsnm{Mannel}, \binits{T.}},
\oauthor{\bsnm{Manohar}, \binits{A.V.}},
\oauthor{\bsnm{Marciano}, \binits{W.J.}},
\oauthor{\bsnm{Masoni}, \binits{A.}},
\oauthor{\bsnm{Matthews}, \binits{J.}},
\oauthor{\bsnm{Meißner}, \binits{U.-G.}},
\oauthor{\bsnm{Melzer-Pellmann}, \binits{I.-A.}},
\oauthor{\bsnm{Mikhasenko}, \binits{M.}},
\oauthor{\bsnm{Miller}, \binits{D.J.}},
\oauthor{\bsnm{Milstead}, \binits{D.}},
\oauthor{\bsnm{Mitchell}, \binits{R.E.}},
\oauthor{\bsnm{Mönig}, \binits{K.}},
\oauthor{\bsnm{Molaro}, \binits{P.}},
\oauthor{\bsnm{Moortgat}, \binits{F.}},
\oauthor{\bsnm{Moskovic}, \binits{M.}},
\oauthor{\bsnm{Nakamura}, \binits{K.}},
\oauthor{\bsnm{Narain}, \binits{M.}},
\oauthor{\bsnm{Nason}, \binits{P.}},
\oauthor{\bsnm{Navas}, \binits{S.}},
\oauthor{\bsnm{Nelles}, \binits{A.}},
\oauthor{\bsnm{Neubert}, \binits{M.}},
\oauthor{\bsnm{Nevski}, \binits{P.}},
\oauthor{\bsnm{Nir}, \binits{Y.}},
\oauthor{\bsnm{Olive}, \binits{K.A.}},
\oauthor{\bsnm{Patrignani}, \binits{C.}},
\oauthor{\bsnm{Peacock}, \binits{J.A.}},
\oauthor{\bsnm{Petrov}, \binits{V.A.}},
\oauthor{\bsnm{Pianori}, \binits{E.}},
\oauthor{\bsnm{Pich}, \binits{A.}},
\oauthor{\bsnm{Piepke}, \binits{A.}},
\oauthor{\bsnm{Pietropaolo}, \binits{F.}},
\oauthor{\bsnm{Pomarol}, \binits{A.}},
\oauthor{\bsnm{Pordes}, \binits{S.}},
\oauthor{\bsnm{Profumo}, \binits{S.}},
\oauthor{\bsnm{Quadt}, \binits{A.}},
\oauthor{\bsnm{Rabbertz}, \binits{K.}},
\oauthor{\bsnm{Rademacker}, \binits{J.}},
\oauthor{\bsnm{Raffelt}, \binits{G.}},
\oauthor{\bsnm{Ramsey-Musolf}, \binits{M.}},
\oauthor{\bsnm{Ratcliff}, \binits{B.N.}},
\oauthor{\bsnm{Richardson}, \binits{P.}},
\oauthor{\bsnm{Ringwald}, \binits{A.}},
\oauthor{\bsnm{Robinson}, \binits{D.J.}},
\oauthor{\bsnm{Roesler}, \binits{S.}},
\oauthor{\bsnm{Rolli}, \binits{S.}},
\oauthor{\bsnm{Romaniouk}, \binits{A.}},
\oauthor{\bsnm{Rosenberg}, \binits{L.J.}},
\oauthor{\bsnm{Rosner}, \binits{J.L.}},
\oauthor{\bsnm{Rybka}, \binits{G.}},
\oauthor{\bsnm{Ryskin}, \binits{M.G.}},
\oauthor{\bsnm{Ryutin}, \binits{R.A.}},
\oauthor{\bsnm{Sakai}, \binits{Y.}},
\oauthor{\bsnm{Sarkar}, \binits{S.}},
\oauthor{\bsnm{Sauli}, \binits{F.}},
\oauthor{\bsnm{Schneider}, \binits{O.}},
\oauthor{\bsnm{Schönert}, \binits{S.}},
\oauthor{\bsnm{Scholberg}, \binits{K.}},
\oauthor{\bsnm{Schwartz}, \binits{A.J.}},
\oauthor{\bsnm{Schwiening}, \binits{J.}},
\oauthor{\bsnm{Scott}, \binits{D.}},
\oauthor{\bsnm{Sefkow}, \binits{F.}},
\oauthor{\bsnm{Seljak}, \binits{U.}},
\oauthor{\bsnm{Sharma}, \binits{V.}},
\oauthor{\bsnm{Sharpe}, \binits{S.R.}},
\oauthor{\bsnm{Shiltsev}, \binits{V.}},
\oauthor{\bsnm{Signorelli}, \binits{G.}},
\oauthor{\bsnm{Silari}, \binits{M.}},
\oauthor{\bsnm{Simon}, \binits{F.}},
\oauthor{\bsnm{Sjöstrand}, \binits{T.}},
\oauthor{\bsnm{Skands}, \binits{P.}},
\oauthor{\bsnm{Skwarnicki}, \binits{T.}},
\oauthor{\bsnm{Smoot}, \binits{G.F.}},
\oauthor{\bsnm{Soffer}, \binits{A.}},
\oauthor{\bsnm{Sozzi}, \binits{M.S.}},
\oauthor{\bsnm{Spanier}, \binits{S.}},
\oauthor{\bsnm{Spiering}, \binits{C.}},
\oauthor{\bsnm{Stahl}, \binits{A.}},
\oauthor{\bsnm{Stone}, \binits{S.L.}},
\oauthor{\bsnm{Sumino}, \binits{Y.}},
\oauthor{\bsnm{Syphers}, \binits{M.J.}},
\oauthor{\bsnm{Takahashi}, \binits{F.}},
\oauthor{\bsnm{Tanabashi}, \binits{M.}},
\oauthor{\bsnm{Tanaka}, \binits{J.}},
\oauthor{\bsnm{Taševský}, \binits{M.}},
\oauthor{\bsnm{Terao}, \binits{K.}},
\oauthor{\bsnm{Terashi}, \binits{K.}},
\oauthor{\bsnm{Terning}, \binits{J.}},
\oauthor{\bsnm{Thorne}, \binits{R.S.}},
\oauthor{\bsnm{Titov}, \binits{M.}},
\oauthor{\bsnm{Tkachenko}, \binits{N.P.}},
\oauthor{\bsnm{Tovey}, \binits{D.R.}},
\oauthor{\bsnm{Trabelsi}, \binits{K.}},
\oauthor{\bsnm{Urquijo}, \binits{P.}},
\oauthor{\bsnm{Valencia}, \binits{G.}},
\oauthor{\bparticle{Van~de} \bsnm{Water}, \binits{R.}},
\oauthor{\bsnm{Varelas}, \binits{N.}},
\oauthor{\bsnm{Venanzoni}, \binits{G.}},
\oauthor{\bsnm{Verde}, \binits{L.}},
\oauthor{\bsnm{Vivarelli}, \binits{I.}},
\oauthor{\bsnm{Vogel}, \binits{P.}},
\oauthor{\bsnm{Vogelsang}, \binits{W.}},
\oauthor{\bsnm{Vorobyev}, \binits{V.}},
\oauthor{\bsnm{Wakely}, \binits{S.P.}},
\oauthor{\bsnm{Walkowiak}, \binits{W.}},
\oauthor{\bsnm{Walter}, \binits{C.W.}},
\oauthor{\bsnm{Wands}, \binits{D.}},
\oauthor{\bsnm{Weinberg}, \binits{D.H.}},
\oauthor{\bsnm{Weinberg}, \binits{E.J.}},
\oauthor{\bsnm{Wermes}, \binits{N.}},
\oauthor{\bsnm{White}, \binits{M.}},
\oauthor{\bsnm{Wiencke}, \binits{L.R.}},
\oauthor{\bsnm{Willocq}, \binits{S.}},
\oauthor{\bsnm{Wohl}, \binits{C.G.}},
\oauthor{\bsnm{Woody}, \binits{C.L.}},
\oauthor{\bsnm{Yao}, \binits{W.-M.}},
\oauthor{\bsnm{Yokoyama}, \binits{M.}},
\oauthor{\bsnm{Yoshida}, \binits{R.}},
\oauthor{\bsnm{Zanderighi}, \binits{G.}},
\oauthor{\bsnm{Zeller}, \binits{G.P.}},
\oauthor{\bsnm{Zenin}, \binits{O.V.}},
\oauthor{\bsnm{Zhu}, \binits{R.-Y.}},
\oauthor{\bsnm{Zhu}, \binits{S.-L.}},
\oauthor{\bsnm{Zimmermann}, \binits{F.}},
\oauthor{\bsnm{Zyla}, \binits{P.A.}}:
{Review of Particle Physics}.
Progress of Theoretical and Experimental Physics
\textbf{2022}(8)
(2022)
{\href{https://arxiv.org/abs/https://academic.oup.com/ptep/article-pdf/2022/8/083C01/45434166/ptac097.pdf}{{https://academic.oup.com/ptep/article-pdf/2022/8/083C01/45434166/ptac097.pdf}}}.
\doiurl{10.1093/ptep/ptac097}.
083C01
\end{botherref}
\endbibitem

\bibitem{de_Grijs2014AJa}
\begin{barticle}
\bauthor{\bsnm{{de Grijs}}, \binits{R.}},
\bauthor{\bsnm{{Wicker}}, \binits{J.E.}},
\bauthor{\bsnm{{Bono}}, \binits{G.}}:
\batitle{{Clustering of Local Group Distances: Publication Bias or Correlated
  Measurements? I. The Large Magellanic Cloud}}.
\bjtitle{\aj}
\bvolume{147},
\bfpage{122}
(\byear{2014})
{\href{https://arxiv.org/abs/1403.3141}{{arXiv:1403.3141}}}.
\doiurl{10.1088/0004-6256/147/5/122}
\end{barticle}
\endbibitem

\bibitem{de_Grijs2020ApJS248}
\begin{barticle}
\bauthor{\bsnm{{de Grijs}}, \binits{R.}},
\bauthor{\bsnm{{Bono}}, \binits{G.}}:
\batitle{{Clustering of Local Group Distances: Publication Bias or Correlated
  Measurements? VII. A Distance Framework out to 100 Mpc}}.
\bjtitle{\apjs}
\bvolume{248}(\bissue{1}),
\bfpage{6}
(\byear{2020})
{\href{https://arxiv.org/abs/2004.00114}{{arXiv:2004.00114}}}
{[astro-ph.GA]}.
\doiurl{10.3847/1538-4365/ab8562}
\end{barticle}
\endbibitem

\bibitem{Salvati2016}
\begin{barticle}
\bauthor{\bsnm{{Salvati}}, \binits{L.}},
\bauthor{\bsnm{{Pagano}}, \binits{L.}},
\bauthor{\bsnm{{Consiglio}}, \binits{R.}},
\bauthor{\bsnm{{Melchiorri}}, \binits{A.}}:
\batitle{{Cosmological constraints on the neutron lifetime}}.
\bjtitle{J. Cosmol. Astro-particle Phys.}
\bvolume{2016}(\bissue{3}),
\bfpage{055}
(\byear{2016})
{\href{https://arxiv.org/abs/1507.07243}{{arXiv:1507.07243}}}
{[astro-ph.CO]}.
\doiurl{10.1088/1475-7516/2016/03/055}
\end{barticle}
\endbibitem

\bibitem{Serebrov2005}
\begin{barticle}
\bauthor{\bsnm{{Serebrov}}, \binits{A.}},
\bauthor{\bsnm{{Varlamov}}, \binits{V.}},
\bauthor{\bsnm{{Kharitonov}}, \binits{A.}},
\bauthor{\bsnm{{Fomin}}, \binits{A.}},
\bauthor{\bsnm{{Pokotilovski}}, \binits{Y.}},
\bauthor{\bsnm{{Geltenbort}}, \binits{P.}},
\bauthor{\bsnm{{Butterworth}}, \binits{J.}},
\bauthor{\bsnm{{Krasnoschekova}}, \binits{I.}},
\bauthor{\bsnm{{Lasakov}}, \binits{M.}},
\bauthor{\bsnm{{Tal'daev}}, \binits{R.}}:
\batitle{{Measurement of the neutron lifetime using a gravitational trap and a
  low-temperature Fomblin coating}}.
\bjtitle{Phys. Lett. B}
\bvolume{605}(\bissue{1-2}),
\bfpage{72}--\blpage{78}
(\byear{2005})
{\href{https://arxiv.org/abs/nucl-ex/0408009}{{arXiv:nucl-ex/0408009}}}
{[nucl-ex]}.
\doiurl{10.1016/j.physletb.2004.11.013}
\end{barticle}
\endbibitem

\bibitem{Byrne1996}
\begin{barticle}
\bauthor{\bsnm{{Byrne}}, \binits{J.}},
\bauthor{\bsnm{{Dawber}}, \binits{P.G.}},
\bauthor{\bsnm{{Habeck}}, \binits{C.G.}},
\bauthor{\bsnm{{Smidt}}, \binits{S.J.}},
\bauthor{\bsnm{{Spain}}, \binits{J.A.}},
\bauthor{\bsnm{{Williams}}, \binits{A.P.}}:
\batitle{{A revised value for the neutron lifetime measured using a Penning
  trap}}.
\bjtitle{EPL (europhysics Lett.}
\bvolume{33}(\bissue{3}),
\bfpage{187}--\blpage{192}
(\byear{1996}).
\doiurl{10.1209/epl/i1996-00319-x}
\end{barticle}
\endbibitem

\bibitem{Mampe1993}
\begin{barticle}
\bauthor{\bsnm{{Mampe}}, \binits{W.}},
\bauthor{\bsnm{{Bondarenko}}, \binits{L.N.}},
\bauthor{\bsnm{{Morozov}}, \binits{V.I.}},
\bauthor{\bsnm{{Panin}}, \binits{Y.N.}},
\bauthor{\bsnm{{Fomin}}, \binits{A.I.}}:
\batitle{{Measuring neutron lifetime by storing ultracold neutrons and
  detecting inelastically scattered neutrons}}.
\bjtitle{Sov. J. Exp. Theor. Phys. Lett.}
\bvolume{57},
\bfpage{82}
(\byear{1993})
\end{barticle}
\endbibitem

\bibitem{Alfimenkov1990}
\begin{barticle}
\bauthor{\bsnm{{Alfimenkov}}, \binits{V.P.}},
\bauthor{\bsnm{{Varlamov}}, \binits{V.E.}},
\bauthor{\bsnm{{Vasil'Ev}}, \binits{A.V.}},
\bauthor{\bsnm{{Gudkov}}, \binits{V.P.}},
\bauthor{\bsnm{{Lushchikov}}, \binits{V.I.}},
\bauthor{\bsnm{{Nesvizhevski{\v{i}}}}, \binits{V.V.}},
\bauthor{\bsnm{{Serebrov}}, \binits{A.P.}},
\bauthor{\bsnm{{Strelkov}}, \binits{A.V.}},
\bauthor{\bsnm{{Sumbaev}}, \binits{S.O.}},
\bauthor{\bsnm{{Tal'Daev}}, \binits{R.R.}}:
\batitle{{Measurement of neutron lifetime with a gravitational trap for
  ultracold neutrons}}.
\bjtitle{Sov. J. Exp. Theor. Phys. Lett.}
\bvolume{52},
\bfpage{373}
(\byear{1990})
\end{barticle}
\endbibitem

\bibitem{Kossakowski1989}
\begin{barticle}
\bauthor{\bsnm{{Kossakowski}}, \binits{R.}},
\bauthor{\bsnm{{Grivot}}, \binits{P.}},
\bauthor{\bsnm{{Liaud}}, \binits{P.}},
\bauthor{\bsnm{{Schreckenbach}}, \binits{K.}},
\bauthor{\bsnm{{Azuelos}}, \binits{G.}}:
\batitle{{Neutron lifetime measurement with a helium-filled time projection
  chamber}}.
\bjtitle{\nphysa}
\bvolume{503}(\bissue{2}),
\bfpage{473}--\blpage{500}
(\byear{1989}).
\doiurl{10.1016/0375-9474(89)90246-7}
\end{barticle}
\endbibitem

\bibitem{Paul1989}
\begin{barticle}
\bauthor{\bsnm{Paul}, \binits{W.}},
\bauthor{\bsnm{Anton}, \binits{F.}},
\bauthor{\bsnm{Mampe}, \binits{W.}},
\bauthor{\bsnm{Paul}, \binits{L.}},
\bauthor{\bsnm{Paul}, \binits{S.}}:
\batitle{{Measurement of the Neutron Lifetime in a Magnetic Storage Ring}}.
\bjtitle{Z. Phys. C}
\bvolume{45},
\bfpage{25}
(\byear{1989}).
\doiurl{10.1007/BF01556667}
\end{barticle}
\endbibitem

\bibitem{Last1988}
\begin{barticle}
\bauthor{\bsnm{{Last}}, \binits{J.}},
\bauthor{\bsnm{{Arnold}}, \binits{M.}},
\bauthor{\bsnm{{D{\"o}hner}}, \binits{J.}},
\bauthor{\bsnm{{Dubbers}}, \binits{D.}},
\bauthor{\bsnm{{Freedman}}, \binits{S.J.}}:
\batitle{{Pulsed-beam neutron-lifetime measurement}}.
\bjtitle{\prl}
\bvolume{60}(\bissue{11}),
\bfpage{995}--\blpage{998}
(\byear{1988}).
\doiurl{10.1103/PhysRevLett.60.995}
\end{barticle}
\endbibitem

\bibitem{Spivak1988}
\begin{barticle}
\bauthor{\bsnm{Spivak}, \binits{P.E.}}:
\batitle{{Neutron Lifetime from Atomic-Energy-Institute experiment}}.
\bjtitle{Sov. Phys. JETP}
\bvolume{67},
\bfpage{1735}--\blpage{1740}
(\byear{1988})
\end{barticle}
\endbibitem

\bibitem{Kosvintsev1986}
\begin{barticle}
\bauthor{\bsnm{{Kosvintsev}}, \binits{Y.Y.}},
\bauthor{\bsnm{{Morozov}}, \binits{V.I.}},
\bauthor{\bsnm{{Terekhov}}, \binits{G.I.}}:
\batitle{{Measurement of neutron lifetime through storage of ultracold
  neutrons}}.
\bjtitle{Sov. J. Exp. Theor. Phys. Lett.}
\bvolume{44},
\bfpage{571}
(\byear{1986})
\end{barticle}
\endbibitem

\bibitem{kosvintsev1980}
\begin{barticle}
\bauthor{\bsnm{Kosvintsev}, \binits{Y.Y.}},
\bauthor{\bsnm{Kushnir}, \binits{Y.A.}},
\bauthor{\bsnm{Morozov}, \binits{V.I.}},
\bauthor{\bsnm{Terekhov}, \binits{G.I.}}:
\batitle{{APPLICATION OF ULTRACOLD NEUTRONS FOR NEUTRON LIFETIME MEASUREMENT.
  (IN RUSSIAN)}}.
\bjtitle{JETP Lett.}
\bvolume{31},
\bfpage{236}
(\byear{1980})
\end{barticle}
\endbibitem

\bibitem{Christensen1972}
\begin{barticle}
\bauthor{\bsnm{{Christensen}}, \binits{C.J.}},
\bauthor{\bsnm{{Nielsen}}, \binits{A.}},
\bauthor{\bsnm{{Bahnsen}}, \binits{A.}},
\bauthor{\bsnm{{Brown}}, \binits{W.K.}},
\bauthor{\bsnm{{Rustad}}, \binits{B.M.}}:
\batitle{{Free-Neutron Beta-Decay Half-Life}}.
\bjtitle{\prd}
\bvolume{5}(\bissue{7}),
\bfpage{1628}--\blpage{1640}
(\byear{1972}).
\doiurl{10.1103/PhysRevD.5.1628}
\end{barticle}
\endbibitem

\bibitem{Crandall2015b}
\begin{barticle}
\bauthor{\bsnm{{Crandall}}, \binits{S.}},
\bauthor{\bsnm{{Houston}}, \binits{S.}},
\bauthor{\bsnm{{Ratra}}, \binits{B.}}:
\batitle{{Non-Gaussian error distribution of 7Li abundance measurements}}.
\bjtitle{Mod. Phys. Lett. A}
\bvolume{30},
\bfpage{1550123}
(\byear{2015})
{\href{https://arxiv.org/abs/1409.7332}{{arXiv:1409.7332}}}.
\doiurl{10.1142/S0217732315501230}
\end{barticle}
\endbibitem

\bibitem{Erler2020}
\begin{barticle}
\bauthor{\bsnm{{Erler}}, \binits{J.}},
\bauthor{\bsnm{{Ferro-Hern{\'a}ndez}}, \binits{R.}}:
\batitle{{Alternative to the application of PDG scale factors}}.
\bjtitle{European Physical Journal C}
\bvolume{80}(\bissue{6}),
\bfpage{541}
(\byear{2020})
{\href{https://arxiv.org/abs/2004.01219}{{arXiv:2004.01219}}}
{[physics.data-an]}.
\doiurl{10.1140/epjc/s10052-020-8115-3}
\end{barticle}
\endbibitem

\bibitem{Chen2003b}
\begin{barticle}
\bauthor{\bsnm{{Chen}}, \binits{G.}},
\bauthor{\bsnm{{Gott}}, \binits{I.} \bsuffix{J.~Richard}},
\bauthor{\bsnm{{Ratra}}, \binits{B.}}:
\batitle{{Non-Gaussian Error Distribution of Hubble Constant Measurements}}.
\bjtitle{\pasp}
\bvolume{115}(\bissue{813}),
\bfpage{1269}--\blpage{1279}
(\byear{2003})
{\href{https://arxiv.org/abs/astro-ph/0308099}{{arXiv:astro-ph/0308099}}}
{[astro-ph]}.
\doiurl{10.1086/379219}
\end{barticle}
\endbibitem

\bibitem{Singh2016}
\begin{barticle}
\bauthor{\bsnm{{Singh}}, \binits{M.}},
\bauthor{\bsnm{{Gupta}}, \binits{S.}},
\bauthor{\bsnm{{Pandey}}, \binits{A.}},
\bauthor{\bsnm{{Sharma}}, \binits{S.}}:
\batitle{{Measurement of Hubble constant: non-Gaussian errors in HST Key
  Project data}}.
\bjtitle{JCAP}
\bvolume{8},
\bfpage{026}
(\byear{2016})
{\href{https://arxiv.org/abs/1506.06212}{{arXiv:1506.06212}}}.
\doiurl{10.1088/1475-7516/2016/08/026}
\end{barticle}
\endbibitem

\bibitem{Huber1981}
\begin{bbook}
\bauthor{\bsnm{Huber}, \binits{P.}}:
\bbtitle{Robust Statistics}.
\bpublisher{Wiley, New York}, \blocation{ }
(\byear{1981})
\end{bbook}
\endbibitem

\bibitem{Efron1994}
\begin{bbook}
\bauthor{\bsnm{Efron}, \binits{B.}},
\bauthor{\bsnm{Tibshirani}, \binits{R.}}:
\bbtitle{An Introduction to the Bootstrap}.
\bpublisher{Chapman and Hall}, \blocation{ }
(\byear{1994})
\end{bbook}
\endbibitem

\bibitem{Conover1999}
\begin{bbook}
\bauthor{\bsnm{Conover}, \binits{W.J.}}:
\bbtitle{Practical Nonparametric Statistics}
vol. \bseriesno{350}.
\bpublisher{john wiley \& sons}, \blocation{ }
(\byear{1999})
\end{bbook}
\endbibitem

\bibitem{Barlow2003}
\begin{botherref}
\oauthor{\bsnm{{Barlow}}, \binits{R.}}:
{Asymmetric Systematic Errors}.
arXiv e-prints,
0306138
(2003)
{\href{https://arxiv.org/abs/physics/0306138}{{arXiv:physics/0306138}}}
{[physics.data-an]}
\end{botherref}
\endbibitem

\bibitem{Barlow2004}
\begin{botherref}
\oauthor{\bsnm{{Barlow}}, \binits{R.}}:
{Asymmetric Statistical Errors}.
arXiv e-prints,
0406120
(2004)
{\href{https://arxiv.org/abs/physics/0406120}{{arXiv:physics/0406120}}}
{[physics.data-an]}
\end{botherref}
\endbibitem

\bibitem{Lista2017}
\begin{bbook}
\bauthor{\bsnm{Lista}, \binits{L.}}:
\bbtitle{{Statistical Methods for Data Analysis in Particle Physics}}
vol. \bseriesno{941}.
\bpublisher{Springer}, \blocation{ }
(\byear{2017})
\end{bbook}
\endbibitem

\bibitem{Audi2017}
\begin{barticle}
\bauthor{\bsnm{{Audi}}, \binits{G.}},
\bauthor{\bsnm{{Kondev}}, \binits{F.G.}},
\bauthor{\bsnm{{Wang}}, \binits{M.}},
\bauthor{\bsnm{{Huang}}, \binits{W.J.}},
\bauthor{\bsnm{{Naimi}}, \binits{S.}}:
\batitle{{The NUBASE2016 evaluation of nuclear properties}}.
\bjtitle{Chinese Physics C}
\bvolume{41}(\bissue{3}),
\bfpage{030001}
(\byear{2017}).
\doiurl{10.1088/1674-1137/41/3/030001}
\end{barticle}
\endbibitem

\bibitem{Possolo2019}
\begin{barticle}
\bauthor{\bsnm{{Possolo}}, \binits{A.}},
\bauthor{\bsnm{{Merkatas}}, \binits{C.}},
\bauthor{\bsnm{{Bodnar}}, \binits{O.}}:
\batitle{{Asymmetrical uncertainties}}.
\bjtitle{Metrologia}
\bvolume{56}(\bissue{4}),
\bfpage{045009}
(\byear{2019}).
\doiurl{10.1088/1681-7575/ab2a8d}
\end{barticle}
\endbibitem

\bibitem{Audi2012}
\begin{barticle}
\bauthor{\bsnm{{Audi}}, \binits{G.}},
\bauthor{\bsnm{{M.}}, \binits{W.}},
\bauthor{\bsnm{{A.~H.}}, \binits{W.}},
\bauthor{\bsnm{{F.~G.}}, \binits{K.}},
\bauthor{\bsnm{{MacCormick}}, \binits{M.}},
\bauthor{\bsnm{{Xu}}, \binits{X.}},
\bauthor{\bsnm{{Pfeiffer}}, \binits{B.}}:
\batitle{{The Ame2012 atomic mass evaluation}}.
\bjtitle{Chinese Physics C}
\bvolume{36}(\bissue{12}),
\bfpage{002}
(\byear{2012}).
\doiurl{10.1088/1674-1137/36/12/002}
\end{barticle}
\endbibitem

\bibitem{Barlow2019}
\begin{botherref}
\oauthor{\bsnm{{Barlow}}, \binits{R.J.}}:
{Practical Statistics for Particle Physics}.
arXiv e-prints,
1905--12362
(2019)
{\href{https://arxiv.org/abs/1905.12362}{{arXiv:1905.12362}}}
{[physics.data-an]}
\end{botherref}
\endbibitem

\bibitem{Podariu2001}
\begin{barticle}
\bauthor{\bsnm{{Podariu}}, \binits{S.}},
\bauthor{\bsnm{{Souradeep}}, \binits{T.}},
\bauthor{\bsnm{{Gott}}, \binits{J.R.} \bsuffix{III}},
\bauthor{\bsnm{{Ratra}}, \binits{B.}},
\bauthor{\bsnm{{Vogeley}}, \binits{M.S.}}:
\batitle{{Binned Cosmic Microwave Background Anisotropy Power Spectra: Peak
  Location}}.
\bjtitle{\apj}
\bvolume{559},
\bfpage{9}--\blpage{22}
(\byear{2001})
{\href{https://arxiv.org/abs/astro-ph/0102264}{{astro-ph/0102264}}}.
\doiurl{10.1086/322409}
\end{barticle}
\endbibitem

\bibitem{Farooq2013PLB}
\begin{barticle}
\bauthor{\bsnm{{Farooq}}, \binits{O.}},
\bauthor{\bsnm{{Crandall}}, \binits{S.}},
\bauthor{\bsnm{{Ratra}}, \binits{B.}}:
\batitle{{Binned Hubble parameter measurements and the cosmological
  deceleration-acceleration transition}}.
\bjtitle{Physics Letters B}
\bvolume{726}(\bissue{1-3}),
\bfpage{72}--\blpage{82}
(\byear{2013})
{\href{https://arxiv.org/abs/1305.1957}{{arXiv:1305.1957}}}
{[astro-ph.CO]}.
\doiurl{10.1016/j.physletb.2013.08.078}
\end{barticle}
\endbibitem

\bibitem{Crandall2014}
\begin{barticle}
\bauthor{\bsnm{{Crandall}}, \binits{S.}},
\bauthor{\bsnm{{Ratra}}, \binits{B.}}:
\batitle{{Median statistics cosmological parameter values}}.
\bjtitle{Phys. Lett. B}
\bvolume{732},
\bfpage{330}--\blpage{334}
(\byear{2014})
{\href{https://arxiv.org/abs/1311.0840}{{arXiv:1311.0840}}}.
\doiurl{10.1016/j.physletb.2014.03.059}
\end{barticle}
\endbibitem

\bibitem{Cowan2019}
\begin{barticle}
\bauthor{\bsnm{{Cowan}}, \binits{G.}}:
\batitle{{Statistical models with uncertain error parameters}}.
\bjtitle{European Physical Journal C}
\bvolume{79}(\bissue{2}),
\bfpage{133}
(\byear{2019})
{\href{https://arxiv.org/abs/1809.05778}{{arXiv:1809.05778}}}
{[physics.data-an]}.
\doiurl{10.1140/epjc/s10052-019-6644-4}
\end{barticle}
\endbibitem

\end{thebibliography}


\end{document}